\documentclass[preprint,12pt]{elsarticle}

\usepackage{listings}    
\usepackage{graphicx}    
\usepackage{subcaption}  
\usepackage{algorithm}   
\usepackage{algorithmic} 
\usepackage{booktabs}    
\usepackage{hyperref}    
\usepackage[htt]{hyphenat} 
\usepackage{microtype} 
\DisableLigatures{}    

\usepackage{caption}
\usepackage{setspace}
\usepackage{multirow}
\usepackage{enumerate}
\usepackage{pdflscape}
\usepackage{pgfplots} 
\usepackage{array}

\usepackage{xcolor}
\definecolor{codegreen}{rgb}{0.25,0.5,0.35}
\definecolor{codegray}{rgb}{0.5,0.5,0.5}
\definecolor{codepurple}{rgb}{0.6,0,0}
\definecolor{backcolour}{rgb}{0.95,0.95,0.92}
\definecolor{colorstring}{rgb}{0.5,0,0.35}
\definecolor{rltred}{rgb}{0.5,0,0}
\definecolor{rltgreen}{rgb}{0,0.5,0}
\definecolor{rltblue}{rgb}{0,0,0.5}
\definecolor{DarkGreen}{rgb}{0.00,0.60,0.00}
\definecolor{ScarletRed}{rgb}{0.80,0.00,0.00}
\definecolor{blizzardblue}{rgb}{0.67, 0.9, 0.93}
\definecolor{green-yellow}{rgb}{0.68, 1.0, 0.18}
\definecolor{dkgreen}{rgb}{0,0.6,0}
\definecolor{gray}{rgb}{0.5,0.5,0.5}
\definecolor{mauve}{rgb}{0.58,0,0.82}
\definecolor{lightgrey}{rgb}{0.90,0.90,0.90}
\definecolor{grey}{gray}{0.75}
\definecolor{light-gray}{gray}{0.80}

\usepackage{xspace}

\usepackage{boxedminipage}
\newenvironment{result}%
{\smallskip
	\noindent
	\let\emph=\textbf
	\begin{boxedminipage}{\columnwidth}\begin{center}\em}%
		{\end{center}\end{boxedminipage}%
}

\usepackage{ifthen}
\newboolean{showcomments}
\setboolean{showcomments}{true} 

\ifthenelse{\boolean{showcomments}}{
	\newcommand{\nbc}[3]{
		{\colorbox{#3}{\bfseries\sffamily\scriptsize\textcolor{white}{#1}}}
		{\textcolor{#3}{\sf\small$\langle$\textit{#2}$\rangle$}}}
	
}{
	\newcommand{\nbc}[3]{}

}

\usepackage{longtable}
\usepackage{xtab}
\newcommand{\totalPapersIncluded}{{$259$}\xspace}

\newcommand{\nrER}{{$9$}\xspace}
\newcommand{\refER}{{\cite{P001} \cite{P063} \cite{P065} \cite{P093} \cite{P137} \cite{P142} \cite{P153} \cite{P170} \cite{P210}}\xspace}

\newcommand{\nAEpapers}{{$181$}\xspace}
\newcommand{\onlyArtificial}{{$25$}\xspace}
\newcommand{\refOnlyArtificial}{{\cite{P011} \cite{P016} \cite{P030} \cite{P045} \cite{P060} \cite{P061} \cite{P073} \cite{P078} \cite{P079} \cite{P083} \cite{P088} \cite{P092} \cite{P128} \cite{P220} \cite{P227} \cite{P240} \cite{P248} \cite{P249} \cite{P252} \cite{P266} \cite{P273} \cite{P285} \cite{P292} \cite{P296} \cite{P308}}\xspace}
\newcommand{\onlineOutliers}{{$3$}\xspace}
\newcommand{\refOnlineOutliers}{{\cite{P006} \cite{P217} \cite{P226}}\xspace}
\newcommand{\nrSutIndustrial}{{$31$}\xspace}
\newcommand{\maxArtificialValue}{{$20$}\xspace}
\newcommand{\maxArtificialRef}{{\cite{P280}}\xspace}
\newcommand{\maxOpenSourceValue}{{$30$}\xspace}
\newcommand{\maxOpenSourceRef}{{\cite{P281}}\xspace}
\newcommand{\maxOnlineValue}{{$6000$}\xspace}
\newcommand{\maxOnlineRef}{{\cite{P217}}\xspace}
\newcommand{\maxIndustrialValue}{{$20$}\xspace}
\newcommand{\maxIndustrialRef}{{\cite{P268}}\xspace}

\newcommand{\totalPapersWithTool}{{$210$}\xspace}
\newcommand{\totalPapersUnnamedTool}{{$71$}\xspace}

\newcommand{\openSourceTools}{{$73$}\xspace}
\newcommand{\notOpenSourceTools}{{$121$}\xspace}
\newcommand{\ratioYesOpenSource}{{$36.3$}\xspace}

\newcommand{\numberPapersAE}{{$181$}\xspace}
\newcommand{\numberPapersWithComparedTools}{{$55$}\xspace}

\newcommand{\totalHEpapers}{{$34$}\xspace}
\newcommand{\totalNumberAcademicParticipants}{{$419$}\xspace}
\newcommand{\totalNumberPractitionerParticipants}{{$503$}\xspace}

\newcommand{\countWhiteBoxTesting}{{$22$}\xspace} 
\newcommand{\countBlackBoxTesting}{{$171$}\xspace}

\newcommand{\selenium}{{$201$}\xspace} 
\newcommand{\cypress}{{$23$}\xspace} 
\newcommand{\puppeteer}{{10}\xspace} 
\newcommand{\playwright}{{10}\xspace} 

\newcommand{\countWithAI}{{$81$}\xspace}

\newcommand{\nrIndustrialCollaborators}{{$46$}\xspace}
\newcommand{\nrHEIndustrial}{{$5$}\xspace}
\newcommand{\refHEIndustrial}{{\cite{P009} \cite{P052} \cite{P103} \cite{P241} \cite{P294}}\xspace}
\newcommand{\percentageIndustrial}{{$17.8$}\xspace}

\newcommand{\papersWithFaults}{{$68$}\xspace}
\newcommand{\papersNotSpecifiedNrFaults}{{$26$}\xspace}

\newcommand{\rqA}{{How many scientific articles were published each year between 2014-2025?}\xspace}
\newcommand{\rqB}{{Which are the most common venues where these articles are published?}\xspace}

\newcommand{\rqC}{{What are the main scientific topics of focus in these articles?}\xspace}
\newcommand{\rqD}{{What kind of scientific contributions do these articles provide?}\xspace}

\newcommand{\rqE}{{What kind of SUTs are used to perform experiments?}\xspace}
\newcommand{\rqF}{{In human studies, what is the number of participants involved in the experiments?}\xspace}

\newcommand{\rqG} {{Which tools are introduced or improved?}\xspace}
\newcommand{\rqH}{{How often tools are open-source?}\xspace}
\newcommand{\rqI} {{In algorithm experiments, how often are proposed tools compared with other tools?}\xspace}
\newcommand{\rqJ}{{In case of algorithm experiments, how many faults are found?}\xspace}

\newcommand{\rqK}{{Which test generation technique is  used most?}\xspace}
\newcommand{\rqL}{{Which existing frameworks and/or libraries are used to enable GUI testing?}\xspace}

\newcommand{\rqM}{{How many studies involve the use of AI techniques?}\xspace}

\newcommand{\rqN}{{How many studies involve collaborations with industry?}\xspace}

\newcommand{\rqAn}{{RQ1}\xspace}
\newcommand{\rqBn}{{RQ2}\xspace}
\newcommand{\rqCn}{{RQ3}\xspace}
\newcommand{\rqDn}{{RQ4}\xspace}
\newcommand{\rqEn}{{RQ5}\xspace}
\newcommand{\rqFn}{{RQ6}\xspace}
\newcommand{\rqGn}{{RQ7}\xspace}
\newcommand{\rqHn}{{RQ8}\xspace}
\newcommand{\rqIn}{{RQ9}\xspace}
\newcommand{\rqJn}{{RQ10}\xspace}
\newcommand{\rqKn}{{RQ11}\xspace}
\newcommand{\rqLn}{{RQ12}\xspace}
\newcommand{\rqMn}{{RQ13}\xspace}
\newcommand{\rqNn}{{RQ14}\xspace}

\journal{Journal of Systems and Software}

\begin{document}

\begin{frontmatter}



\title{A Survey on Web Testing: On the Rise of AI and Applications in Industry}

 \author[label1]{Iva Kertusha}
 \author[label1]{Gebremariam Assres}
 \author[label2]{Onur Duman}
 \author[label1]{Andrea Arcuri}
 \affiliation[label1]{organization={Kristiania University of Applied Sciences},
             addressline={Kirkegata 24-26},
             city={Oslo},
             postcode={0153},
             country={Norway}}

 \affiliation[label2]{organization={Glasgow Caledonian University},
             addressline={Cowcaddens Road},
             city={Glasgow},
             postcode={G4 0BA},
             state={Scotland},
             country={UK}}

\begin{abstract}

Web application testing is an essential practice to ensure the reliability, security, and performance of web systems in an increasingly digital world.
This paper presents a systematic literature survey focusing on web testing methodologies, tools, and trends from 2014 to 2025. By analyzing \totalPapersIncluded research papers, the survey identifies key trends, demographics, contributions, tools, challenges, and innovations in this domain. In addition, the survey  analyzes the experimental setups adopted by the studies, including the number of participants involved and the outcomes of the experiments. 
Our results show that web testing research has been highly active, with ICST as the leading venue. Most studies focus on novel techniques, emphasizing automation in black-box testing. Selenium is the most widely used tool, while industrial adoption and human studies remain comparatively limited.
The findings provide a detailed overview of trends, advancements, and challenges in web testing research,  the evolution of automated testing methods, the role of artificial intelligence in test case generation, and gaps in current research. Special attention was given to the level of collaboration and engagement with the industry. A positive trend in using industrial systems is observed, though many tools lack open-source availability.

\end{abstract}
\begin{highlights}
	\item Analyze Web Testing methodologies, tools and trends on the last decade (2014-2025)
	\item Analyze the rate of collaboration with industry.
	\item Analyze the role of AI in web testing.
\end{highlights}

\begin{keyword}
Web Testing \sep GUI Testing  \sep front-end  \sep survey  \sep SBST  \sep AI \sep  fuzzing

\end{keyword}

\end{frontmatter}


\section{Introduction}
The World Wide Web, commonly known as the Web, is the largest internet-based application and has experienced remarkable growth since its public debut in 1993~\cite{gandon2022never}.
In an annual  report published by Verisign, a global provider of critical internet infrastructure and domain name registry services, it is estimated that there are 354 million existing domains in the first quarter of 2023.\footnote{https://blog.verisign.com/domain-names/verisign-q1-2023-the-domain-name-industry-brief/}
In 2024, Statista, a global data and business intelligence platform,  reports 5.44 billion internet users worldwide.\footnote{https://www.statista.com/topics/1145/internet-usage-worldwide/}
SimilarWeb, a company specialized on web analytics,\footnote{https://www.similarweb.com/} reports the top three most visited industries in 2024 are Search Engines, News \& Media, and TV Movies \& Streaming~\cite{xavier2024web}.
This highlights the significance of the web frontend, as it directly influences user engagement and interaction across these highly visited industries.

The web frontend plays a crucial role in shaping user experiences, serving as the interface between users and the underlying functionality of a web application~\cite{thielsch2014user,xavier2024web}.
A well-designed frontend enhances usability, accessibility, and engagement by providing intuitive navigation and visually appealing layouts. As the first point of interaction, the frontend significantly impacts user perception and satisfaction of the web application.

Web frontend testing is vital to ensure that web applications provide a seamless and reliable user experience~\cite{S2,S4}.
By validating the functionality, usability, and responsiveness of the user interface, frontend testing helps identify and resolve potential issues before an official release or update. This not only enhances user satisfaction but also upholds the application’s reputation by preventing bugs, inconsistencies, and performance bottlenecks.

Web frontend testing presents several challenges due to the dynamic and diverse nature of modern web applications. The complexity of interactive elements, asynchronous behaviors, and responsiveness pose significant challenges in testing web frontend applications. Maintaining up-to-date test environments and addressing the rapid evolution of web technologies further complicates the testing process.

As web applications become more complex and dynamic, testing approaches must adapt to ensure reliability, usability, and performance. Web development has undergone significant transformations in recent years, introducing new technologies, frameworks, and practices that need a corresponding evolution in testing methodologies. 

In this survey, we aim to identify key trends, advancements, and challenges of the web testing domain over the past decade. To achieve a comprehensive understanding of this field, we have analyzed \totalPapersIncluded papers covering the period 2014-2025. Additionally, we have formulated 14 research questions, that attempt to explore and analyze the advancements, challenges, and trends in web testing since 2014.

The article is organized as follows.
Section~\ref{sec:background} introduces important background information needed to better understand the rest of this article.
Section~\ref{sec:relatedwork} discusses related work.
Section~\ref{sec:ResearchMethod} describes the paper selection process.
In Section~\ref{sec:study} we answer and discuss our research questions.
Our reflections and insight from answering these research questions are presented in Section~\ref{sec:discussion}.
Threats to validity are discussed in Section~\ref{sec:threats}.
Section~\ref{sec:conclusions} provides the conclusion of the article.

\section{Background}
\label{sec:background}

\subsection{ Web Applications}
Since its introduction in 1991~\cite{history_web_cern}, the web has undergone significant evolution, marked by continuous advancements. The web is quite complex, comprising of, but not limited to, frontend technologies, backend technologies, database technologies, protocols, etc.
Initially, all Web documents were simply HTML pages~\cite{history_web_cern}.
The introduction of Cascading Style Sheets (CSS) enhanced the presentation and layout of web pages, while JavaScript enabled dynamic and interactive content~\cite{history_web_cern}. The development of server-side technologies and databases facilitated the creation of complex web applications, expanding the web's functionality beyond static content delivery. In recent years, trends such as Progressive Web Applications (PWAs), responsive design, and the integration of artificial intelligence have shaped web development practices. Web 4.0~\cite{almeida2017concept} is the current phase of internet,  often referred to as the ``Intelligent Web''.  It represents the next phase of internet evolution, integrating artificial intelligence, machine learning, and advanced automation to create highly personalized, context-aware, and interconnected digital experiences~\cite{BG_almeida2017concept}.

\subsection{Javascript and Asynchronousness}
JavaScript is a fundamental technology of modern web development, playing a crucial role in creating dynamic and interactive web applications. Initially developed in 1995 by Netscape, JavaScript was designed to enhance static web pages with simple scripting capabilities~\cite{BG1_javascript_first20years}.
According to Statista, JavaScript is the most popular language of 2024.\footnote{https://www.statista.com/statistics/793628/worldwide-developer-survey-most-used-languages/}
Over time, JavaScript has evolved into a powerful language that operates both on the client and server sides. 
Popular and  powerful JavaScript frameworks include  React, Angular, and Vue.js in frontend development. The rise of Node.js has extended JavaScript beyond the browser, enabling server-side development.  JavaScript is a cornerstone of web development, continuously evolving with new tools, libraries, and performance optimizations to support complex web applications and real-time interactivity.

JavaScript possesses several features that make it appealing to developers, including dynamic typing, support for event-driven programming, asynchronous event handling, first-class functions, and its weakly typed nature. These very same characteristics make JavaScript challenging to test. Its dynamic nature introduces complexities such as unpredictable type conversions, runtime errors, and difficulties in handling asynchronous operations.

\subsection{Web Testing}
\label{sec:background:webtesting}
Testing is an important phase in the software development lifecycle, as it ensures that applications adhere to quality standards, fulfill functional requirements, and meet user expectations. Testing a web application amounts to almost 50\% of the development cost~\cite{ammannintroduction, el2006multi}. Given the complexity of web applications now-a-days, the need for highly efficient and effective testing tools has increased.

Manual testing involves executing test cases manually without the use of automated tools to identify defects in a software application. Testers simulate user interactions to evaluate functionality, usability, and overall system behavior, making it essential for exploratory, usability, and ad-hoc testing scenarios. However manual testing is prone to human errors, limited exploration capabilities and huge costs. Given the rapid development cycles of web applications, it is essential to automate the testing process as much as possible to ensure efficiency and maintain quality.
One of the earliest automated tools for web testing is Selenium,\footnote{https://www.selenium.dev/} which enables controlling a browser programmatically.
Now Selenium has evolved into a suite of tools, including Selenium WebDriver, Selenium Grid and Selenium IDE.  Being open-source, supporting multiple programming languages and its integration with various frameworks make Selenium a popular choice for developers and testers.
Other tools include Puppeteer,\footnote{https://pptr.dev/} Playwright\footnote{https://playwright.dev/} and Cypress.\footnote{https://www.cypress.io/}
It is important to emphasize that these libraries and tools primarily serve to automate the execution of test cases by enabling controlling a browser programmatically, while the creation and design of the test cases remain a manual process carried out by testers.

This led to the need of automated test case generation.
Within the academic literature there are a few tools on automation of test case generation, with some of the earliest tools being ATUSA~\cite{P146}, AutoBlackTest~\cite{P109} and GUITAR~\cite{P163}.
Other popular tools include Crawljax~\cite{P073}, Testar~\cite{P108}, QExplore~\cite{P157}, and WebExplor~\cite{P005}.

There are two main principles based on which most of these tools are built on: \emph{white-box} and \emph{black-box}. White-box testing involves examining the internal structure, code, and logic of a system to ensure correctness. It typically requires knowledge of the software's implementation. In contrast, black-box testing focuses on evaluating a system's functionality based on input and expected output, without considering its internal workings. 

\section{Related Work}
\label{sec:relatedwork}

\begin{table}[htpb]
	\centering
	\caption{
		Existing surveys about web testing.
		We report the year of publication, and how many papers were analyzed.
		When this latter information is not specified, we report in parentheses () the total number of references in such articles.
		We also report the time-period of the analyzed articles, as well as specifying if the articles focus only on some specific aspects of web testing.
	}
	\label{tab:surveys}
	\begin{tabular}{rrrll}
		\toprule
		Year & Reference & Nr. Papers & Period & Focus \\
		\midrule
		2006 & \cite{S1} & (27) & All previous & \\
		2013 & \cite{S2} &   79 & 2000-2011 & \\
		2014 & \cite{S3} & (51) & Previous decade & \\
		2014 & \cite{S4} &  95  & 2000-2013 & \\
		2014 & \cite{S5} & (82) & Previous 2 decades & \\
		2014 & \cite{S19} & (237) & All previous & Extraction GUI Models \\
		2016 & \cite{S6} & (118) & Previous 2 decades & Testing\&Analysis \\
		2016 & \cite{S17} & (24) & All previous & \\
		2016 & \cite{S7} &   45  & 2006-2015 & Model-Based Testing \\
		2018 & \cite{S8} &   79  & 1988-2017 & Regression Test Case Generation \\
		2019 & \cite{S16} & 27 & 2006-2016 & Quality \\
		2021 & \cite{S9} &  142  & 2020 & Grey Literature \\
		2021 & \cite{S10} &   20  & 2009-2021 & Selenium Framework \\
		2021 & \cite{S11} &  12  & Previous decade & Black-Box Fuzzing \\
		2022 & \cite{S12} &  98  & 2008-2021 & \\
		2023 & \cite{S13} &  26  & 2012-2021 & \\
		2023 & \cite{S14} &  66  & 2000-2022 & Testing Web Page Layout \\
		2024 & \cite{S18} & (199) 314 & 2014-2023 & \\
		2025 & \cite{S15} &  72  & 2013-2023 & \\
		\bottomrule
	\end{tabular}
\end{table}

Testing web applications is an important problem of practical value.
As such, throughout the years lot of research work has been carried out on this topic.
Such work has been analyzed and categorized in few surveys in the last 20 years.
Table~\ref{tab:surveys} provides a summary of these existing surveys.

One of the earliest work is from Di Lucca and Fasolino~\cite{S1}, in 2006.
The state-of-the-art of web testing was discussed, although it was not a formal survey or systematic literature review.
In its reference list in~\cite{S1}, 27 articles were cited in total.
Nearly a decade later, in 2014, another two non-systematic surveys analyzed the state-of-the-art in web testing research~\cite{S3,S5}.
A mapping study was carried out in 2013~\cite{S2}, analyzing  79 published articles.
This was followed by a systematic literature review by a subset of the same authors in 2014~\cite{S4}, analyzing 95 articles.
A short (27 articles) survey appeared afterwards in 2019, covering a similar time period, i.e., 2006-2016.
A survey focused on how GUI models (including from web applications) can be extracted for automated testing was presented in 2014~\cite{S19}.
These early surveys~\cite{S1,S2,S3,S4,S5,S19} cover the periods before 2014, which is the starting point for our survey.

In 2016, a position paper~\cite{S6} discussed the latest advances in web testing in the previous two decades, but including as well other kinds of software analyses.
In the same year, a systematic literature review~\cite{S7} analyzed 45 articles, but focusing only on ``Model-Based Testing''.
Another review appeared in 2016, but only included up to 24 articles~\cite{S17}.
In the following years, other surveys focusing on some specific aspects of testing web applications are for example about
``Regression Test Case Generation''~\cite{S8} (79 articles),
``Quality of Tests''~\cite{S16} (27 articles),
``Selenium Framework''~\cite{S10} (20 articles),
``Black-Box Fuzzing''~\cite{S11} (12 articles)
and
``Testing Web Page Layout''~\cite{S14} (66 articles).
Furthermore, besides analyzing the academic literature, there has been also cases like in~\cite{S9} in which the ``Grey Literature'' about web testing has been analyzed (142 sources).

There are three recent peer-reviewed surveys about web testing that are close to our work~\cite{S12,S13,S15}, analyzing from 26 to 98 (but this latter going back to 2008) articles, considering a similar time period to our 2014-2025 (see Table~\ref{tab:surveys}), although not covering the last two years (i.e., 2024-2025).
By analyzing \totalPapersIncluded articles, we can provide a more thorough view of the last decade in web testing research.
Furthermore, with 14 research questions, we do analyze several important aspects of web testing that were not discussed before.
For example, considering the current success stories of artificial intelligence techniques in many different aspects of science and engineering, we make sure to analyze how they have been used in web testing (as we will discuss in Section~\ref{rq:M}).
Also, our survey focus only on testing.
Other aspects like
``automated repair'' of code~\cite{mahajan2021effective}
and
``fault localization''~\cite{ocariza2016automatic}
(which were included for example in~\cite{S15})
are not analyzed here in our survey, as not in scope.

A non-peer-reviewed survey~\cite{S18}, published on arXiv in 2024, covers similar topics compared to our survey.
In~\cite{S18} it is stated that 314 articles are reviewed, but only 199 are cited.
So we cannot check if in our survey we missed any, or those other articles cover different aspects of web applications that we do not consider in our survey.
That survey does not cover the latest 2 years, i.e., the time-period 2024-2025.
Furthermore, with only six research questions~\cite{S18}, there are several important research questions that we study in our survey that are not covered in~\cite{S18}.
These for example include an analysis of the kinds of scientific contributions (\rqDn),
how studies with human subjects were performed (\rqFn),
how novel tools were compared (\rqIn),
which frameworks are most used to enable GUI-frontend testing (\rqLn),
the role of AI-techniques (\rqMn)
and
details of applications in industry (\rqNn).

Compared to existing surveys in the literature, our survey provides the important contribution of a more in-depth analysis of the last 12 years  of research in web frontend testing.

\section{Research Method}
\label{sec:ResearchMethod}
\subsection{Research questions}
To have a better understanding of the evolution and current state of web testing, we formulated the following Research Questions, which we are going to answer in this survey.
\label{sec:researchQuestions}
\begin{itemize}
	\item \rqAn: \rqA
	\item \rqBn: \rqB
	\item \rqCn: \rqC
	\item \rqDn: \rqD
	\item \rqEn: \rqE
	\item \rqFn: \rqF
	\item \rqGn: \rqG
	\item \rqHn: \rqH
	\item \rqIn: \rqI
	\item \rqJn: \rqJ
	\item \rqKn: \rqK
	\item \rqLn: \rqL
	\item \rqMn: \rqM
	\item \rqNn: \rqN
\end{itemize}
\rqAn and \rqBn aim to provide a general overview of the demographics of the publications.
RQ3 and RQ4 focus on identifying the main topics and contributions of the papers, acknowledging that Web Testing is a broad and diverse field with several areas of focus.
In cases where the paper discusses an experiment, we discuss the used SUTs in RQ5. If the experiments involve human participants, RQ6 addresses this aspect.
Through RQs 7,8 and 9, we aim to analyze the tools presented in the studies and understand their lifespan, usage and maintenance.
RQ10 attempts to check the effectiveness of the proposed tools and techniques.
RQs 11, 12 and 13 attempt to better understand the approaches on the proposed methods and techniques.
RQ11 examines the balance between white-box and black-box testing approaches, and RQ12 investigates the extent to which libraries and frameworks are used to facilitate GUI testing.
RQ13 and RQ14 address two particularly important aspects: RQ13 explores the integration of AI techniques in web testing, while RQ14 evaluates the degree of collaboration between academia and industry within the selected studies.
We believe that these research questions  address various key aspects and provide a comprehensive understanding of the current state of the literature.
\subsection{Databases and Search Queries}
\label{sec:DatabasesAndSearchQueries}
To identify all relevant papers, we conducted a comprehensive search across seven major databases: IEEE, ACM, MIT Libraries, Web of Science (WoS), Wiley, ScienceDirect, and SpringerLink. These databases are widely used in the research community and contain peer-reviewed articles from reputable conferences and journals. They have been used in previous surveys in software engineering research, including~\cite{S15, S14,S18}.

Initially, we identified key terms pertinent to our research objectives to ensure the retrieval of relevant papers from each database. Using these keywords, we constructed search queries. Since each database employs a unique search interface and format, we adjusted the queries accordingly. Furthermore, we applied additional filters specific to each database to refine the search results. For example, in MIT Libraries, we utilized filters for publication date, language, and material type (e.g., books, journals, articles). 

\begin{longtable}{p{2cm} p{8cm} >{\raggedright\arraybackslash}p{3cm}}
	\caption{The queries executed for each database and the resulting number of papers from each query.}
	\label{table:dbqueries} \\
	\toprule
	Database & Query & Nr of papers \\
	\midrule
	\endfirsthead
	
	\toprule
	Database & Query & Nr of papers \\
	\midrule
	\endhead
	
	\midrule
	\multicolumn{3}{r}{\emph{Continued on next page}} \\
	\midrule
	\endfoot
	
	\bottomrule
	\endlastfoot
	
	IEEE &
	(("All Metadata": "web NEAR/2 test*")
	OR
	("All Metadata": "web NEAR/2 fuzz*")
	OR
	("All Metadata": "UI NEAR/2 test*")
	OR
	("All Metadata": "UI NEAR/2 fuzz*")
	OR
	("All Metadata": "GUI NEAR/2 test*")
	OR
	("All Metadata": "GUI NEAR/2 fuzz*")
	OR 
	(("Document Title": "GUI") AND (("All Metadata": "test") OR ("All Metadata": testing)) AND ("All Metadata": "web"))
	OR
	(("Document Title": test* OR "Document Title": fuzz*) AND ("Document Title": "web app*")))
	AND NOT(("All Metadata": fuzzy))
	AND (NOT ("Document Title": mobile OR "Document Title": android OR "Document Title": iOS OR "Document Title": desktop)) 
	& phase 1 - 160  phase 2 - 37 \\
	
	ACM &
	Title: (test*) AND (NOT (Title:(mobile) OR Title: (android)))
	AND ((Abstract:(GUI) OR Abstract:(front-end) OR Abstract:(frontend) OR Abstract:(front end))
	AND (Abstract:(test*) OR Abstract:(fuzz*)) AND (Abstract: (Web)))
	& phase 1 - 58  phase 2 - 5 \\
	
	ScienceDirect &
	Title, abstract, keywords: 
	("test" OR "testing" OR "fuzz" OR "fuzzing") AND ("user interface" OR "UI" OR "GUI") AND ("web application" OR "web")
	& phase 1 - 57  phase 2 - 32 \\
	
	Wiley &
	"test* OR fuzz*" in Abstract and "GUI OR frontend OR front-end OR front end OR user interface OR UI" in Abstract and "test*" in Title
	& phase 1 - 13  phase 2 - 0 \\
	
	Web of Science &
	(((TI=("test*" OR "testing*" OR "fuzz*" OR "fuzzing*")) 
	AND AB=("test*" OR "testing*" OR "fuzz*" OR "fuzzing*")) 
	AND AB=("web")) 
	AND AB=("GUI" OR "graphical user interface" OR "UI" OR "user interface" OR "front end" OR "front-end" OR "frontend")
	& phase 1 - 61  phase 2 - 6 \\
	
	SpringerLink &
	("test" OR "testing" OR "fuzz" OR "fuzzing") AND ("web application") AND ("GUI" OR "graphical user interface" OR "UI" OR "user interface" OR "front end" OR "front-end" OR "frontend")
	& phase 1 - 236  phase 2 - 161 \\
	
	MIT &
	Title containing these search terms: testing OR test OR fuzz OR fuzzing
	
	AND
	
	Abstract/description containing these search terms: "test" OR "testing" OR "fuzz" OR "fuzzing"
	
	AND
	
	Abstract/description containing these search terms: "GUI" OR "front end" OR "front-end" OR "UI" OR "user interface"
	
	AND
	
	Abstract/description containing these search terms: "web application"
	& phase 1 - 0  phase 2 - 20 \\
	
	\midrule
	Total & & phase 1 - 625   phase 2 - 261 \\
	
\end{longtable}

Table ~\ref{table:dbqueries} shows the queries we run in every database.

\subsection{Paper Selection Criteria}
\label{sec:PaperSelectionCriteria}
Our initial search, conducted in 17-24/04/2024 resulted in 625 papers. This large number can be attributed to the fact that our main keywords are quite common and appear even in papers that might not be related to web testing. We decided to eliminate the papers that did not fit the criteria below.

\begin{itemize}
	\item Paper is in English.
	\item Paper is published between 2014-2024.
	\item Paper is related to web testing.
	\item Paper is not a thesis or a patent.
	\item Paper is not a survey or a systematic literature review
\end{itemize}
The first two authors went over all the papers independently, decided on the list of papers to be included, and then compared their respective lists. In case of any discrepency, the issue was solved by the fourth author. After checking all the papers retrieved initally from the databases, we ended up with 145 papers. These papers were reviewed again from the authors, and another 21 papers were excluded.

The second phase in paper collection involves snowballing. We followed the guidelines described in~\cite{wohlin2014guidelines}. We conducted forward and backward snowballing, and we ended up adding 90 new papers. Our final list amounts to 214 papers.
Based on this selection of articles, a first draft of this survey was published on arXiv~\cite{kertusha2025survey}.

To get a better picture of the current state-of-the-art in web testing,
in particular to take into account the latest development in AI-based testing,
we went through a third phase during
12-15/05/2025,
where we added and analyzed papers published in 2024 and 2025. We executed again the steps described in phases one and two,  this time setting the publication years 2024 and 2025.
After the third phase, the total number of papers included in this study is \totalPapersIncluded.

\begin{figure}[t!]
	\centering
	\includegraphics[width=\textwidth]{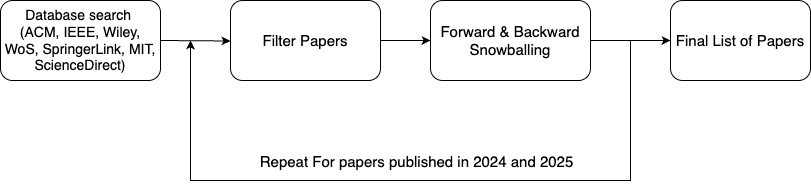}
	\caption{Paper selection process} 
\end{figure}
\subsection{Data Extraction}
\label{sec:DataExtraction}
The study comprises an extensive collection of papers. To manage and analyze the data, a structured Google spreadsheet was created. This document contains the list of selected papers alongside the extracted data, which is organized to address the research questions outlined in Section~\ref{sec:researchQuestions}. Given the extensive collection of papers analyzed in this study, the dataset was divided between the first two authors for initial review and data extraction. Upon completion of the spreadsheet containing all extracted data, the results were collaboratively reviewed and thoroughly cross-verified by the remaining two authors to ensure accuracy and consistency.

\section{Empirical Study}
\label{sec:study}
\subsection{\rqAn - \rqA}
\label{rq:A}
\begin{figure}[t!]
	\centering
	\includegraphics[width=0.6\textwidth]{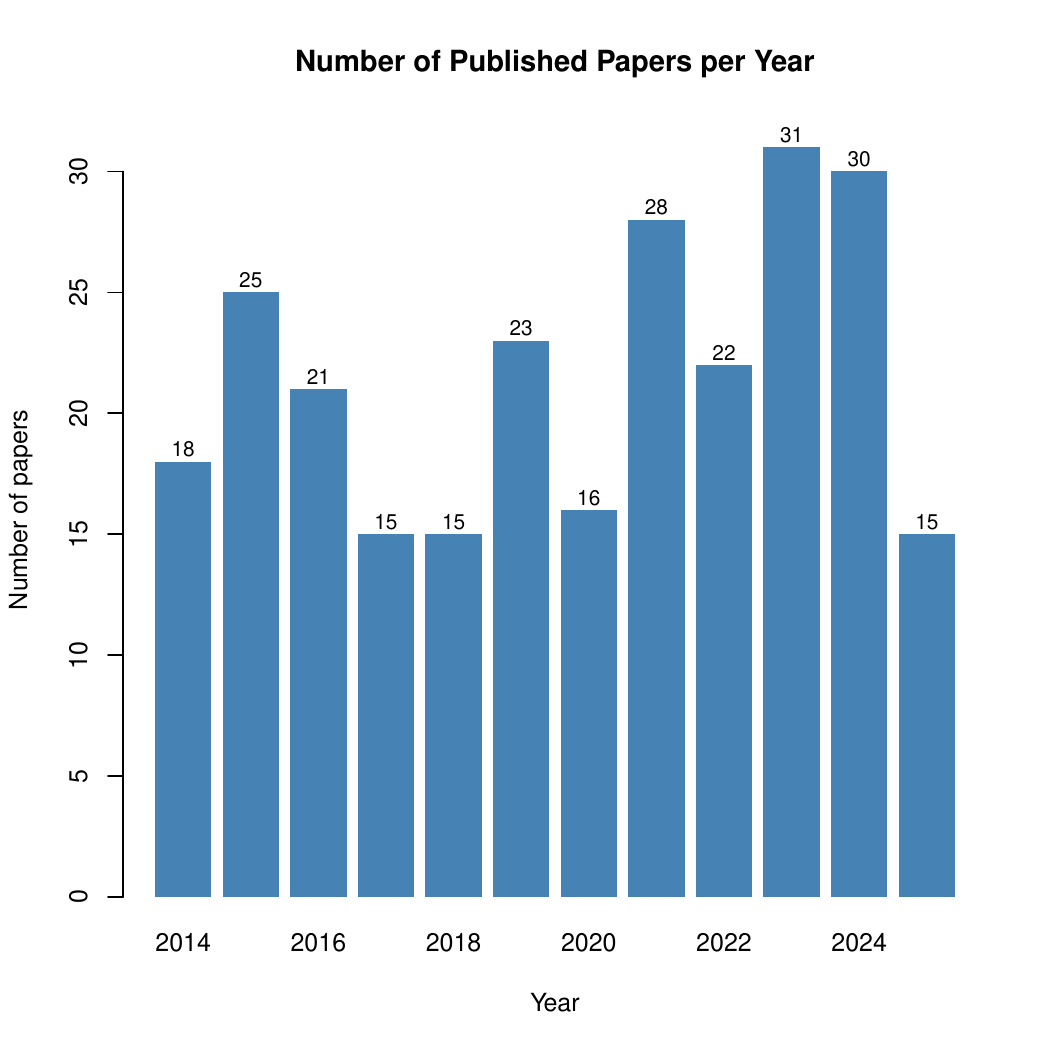}
	\caption{Number of published papers per year} 
	\label{fig:fig_rq1} 
\end{figure}
Figure~\ref{fig:fig_rq1} illustrates the publication rate of papers on web testing from January 2014 to May 2025.
The publication rate has remained relatively stable, fluctuating between 15 and 31 papers per year, with the peak reached in 2023.
In 2025 up to May, 15 papers have already been published, indicating a promising trend that the rate of publication is likely to continue without decline.

\begin{result}
	{\bf RQ1:} The field of research on Web Testing has been significantly active in the last decade, with no significant signs of any slow down.
\end{result}

\subsection{\rqBn - \rqB}
\label{rq:B}

Table~\ref{tab:rq2} shows all the venues in which the different articles have been published.
These venues are sorted based on the number of published articles.
For each venue, it is indicated if it is a journal, a conference or a workshop.
Due to the large number of different venues (more than 100), we only report the acronym of the venues in which at lest
two articles have been published.
All others are grouped together into the ``Other conferences'' and ``Other journals'' categories.

From this list, the most common venue is ICST: the IEEE International Conference on  Software Testing, Verification and Validation.
Furthermore, the 4th most common venue is ICSTW, which is the set of workshops run at ICST.
Arguably, ICST is among the most prestigious conferences in software testing.
The 2nd and 3rd common venues are FSE (ACM International Conference on the Foundations of Software Engineering, previously known as ESEC/FSE) and ICSE (IEEE/ACM International Conference on Software Engineering).
These are arguably the most important conferences in software engineering, in general (and not specific to testing).

Regarding journals, the most commons are
JSS (Journal of Systems and Software),
STVR (Software Testing, Verification and Reliability),
SQJ (Software Quality Journal),
TSE (IEEE Transactions on Software Engineering),
EMSE (Empirical Software Engineering),
IST (Information and Software Technology)
and
TOSEM (ACM Transactions on Software Engineering and Methodology).
These are arguably the most important journals focused on software engineering and software testing.

This data shows a clear importance of web testing and the major efforts of the research community in addressing its research challenges.

Compared to other surveys covering the same period like~\cite{S15},
we see similar trends when it comes to journals.
However, our data is significantly different when it comes to conferences.
For example, in~\cite{S15} the most common venue ICST (and its workshops ICSTW) is not even mentioned.

\begin{result}
	{\bf \rqBn:} ICST is the most common venue where to find the latest advances in Web Testing research, followed by FSE and ICSE.
	The top journals in software engineering all share a similar amount of articles on Web Testing.
\end{result}

\subsection{\rqCn - \rqC}
\label{rq:C}

Testing web frontend applications present many challenges.
As such, several different articles in the literature addressed different problems in this domain.
The following is qualitative categorization of these addressed topics.
As for all other qualitative categorizations in these articles, it was first proposed by one of us authors after reading all the reviewed articles, and then refined via discussions with all other authors.
Each article's category was reviewed by at least two of us authors.
An article can focus on one or more topics/contributions.
Those categories are as follows.

\begin{description}
	\item[Test Generation:] articles involving the automated generation of new test cases.
	\item[Test Maintenance:] articles that focus on the maintainability of existing test cases. Web applications evolve, and ideally existing test cases should fail only if revealing faults, and not for other reasons.
	\item[Testing Framework:] articles related to how test cases can be (manually) written and/or executed.
	\item[Model Inference:] articles describing how models can be derived from web frontend GUIs. These models can then be used for different purposes, typically for test generation.
	\item[Oracle:] articles presenting or evaluating new techniques to automatically detect faults in web applications.
	\item[Coverage Criteria:] articles presenting new or evaluating existing coverage criteria for web testing.
	\item[Test Prioritization:] and selection, related to regression testing for web applications.
	\item[Taxonomy:] articles presenting new taxonomies related to web testing.
	
	\begin{table}[ht]
\centering
\caption{Categories and Total papers per category} 
\label{tab:rq4_2}
\begin{tabular}{lr}
  \toprule
Category & Nr.Papers \\ 
  \midrule
Test Generation & 124 \\ 
  Test Maintenance & 52 \\ 
  Testing Framework & 49 \\ 
  Model Inference & 35 \\ 
  Oracle & 21 \\ 
  Coverage Criteria & 14 \\ 
  Test Prioritization & 7 \\ 
  Taxonomy & 4 \\ 
  Dataset & 2 \\ 
   \bottomrule
\end{tabular}
\end{table}

	\item[Dataset:] articles presenting and discussing curated datasets to enable and simplify future research studies.
\end{description}

\begin{table}[ht]
\centering
\caption{Maintenance} 
\label{tab:rq4_maintenance}
\begin{tabular}{p{3cm}>{\raggedleft}p{2cm}p{6cm}}
  \toprule
Maintenance & Nr.Papers & List.papers \\ 
  \midrule
locator & 29 & \cite{P003} \cite{P008} \cite{P009} \cite{P012} \cite{P039} \cite{P058} \cite{P089} \cite{P104} \cite{P105} \cite{P115} \cite{P119} \cite{P143} \cite{P179} \cite{P182} \cite{P183} \cite{P184} \cite{P185} \cite{P194} \cite{P195} \cite{P196} \cite{P201} \cite{P204} \cite{P207} \cite{P211} \cite{P253} \cite{P254} \cite{P267} \cite{P268} \cite{P292} \\ 
  other & 8 & \cite{P004} \cite{P071} \cite{P083} \cite{P090} \cite{P199} \cite{P223} \cite{P289} \cite{P294} \\ 
  flaky & 8 & \cite{P064} \cite{P190} \cite{P210} \cite{P212} \cite{P238} \cite{P250} \cite{P287} \cite{P296} \\ 
  page-object & 7 & \cite{P082} \cite{P140} \cite{P198} \cite{P200} \cite{P208} \cite{P232} \cite{P242} \\ 
   \bottomrule
\end{tabular}
\end{table}

Table~\ref{tab:rq4} shows, for each observed combination of contribution, the number articles in that group, with references.
Table~\ref{tab:rq4_2} is based on the same data, but only showing the number of articles for each contribution in isolation.

From this data we can see that, by far, the most common topic is automated test generation.
Given a web application, many techniques have been proposed to automatically generate test cases, using for example libraries/frameworks such as Selenium to control a browser to interact programmatically with the web application.
Different strategies can be designed to best explore the graphical user interface (GUI), where transitions from page to page can be represented with models.
Different tools/techniques build models of the GUI to enable a more efficient test generation.
However, inferred models can be used also for other activities.

The investigation of frameworks/libraries for writing and executing tests is common as well.
Related to generating tests, there has been novel work on how to automatically detect faults in web applications (i.e., oracle problem), as well as to define coverage criteria for web test cases.
Test priorization (and selection), as well as the definition of taxonomies, have received comparatively only little attention.
We found only two  articles dealing with the collection and specification of a dataset for experimentation in web testing~\cite{P031} and ~\cite{P299}.

An important topic of research is test maintenance.
Once a test case is manually written, or automatically generated, how will it be used in future versions of the evolving tested web application?
This topic category can be further clustered in three main groups (\emph{locator}, \emph{page-object} and \emph{flaky}), as shown in Table~\ref{tab:rq4_maintenance}.
Most of the work has been dealing with HTML \emph{locators}.
To interact with a browser, a test case need to specify which buttons to click, and which forms to fill in, for example.
This is done by locating on the HTML page the different widgets that a user can interact with.
Each element in the HTML page needs to be uniquely and un-ambiguously identified and located.
This is typically done with
XPath Locators\footnote{https://www.w3.org/TR/xpath/}
or
CSS Selectors.\footnote{https://www.w3.org/TR/selectors-4/}
A problem, though, is that the frontend of a web application can change significantly (e.g., to improve its design), without changing its functionalities.
As such, HTML element locators can easily become outdated, leading the test cases using them to fail, although no semantic change was applied in the web application.
This is a major problem in regression testing.
A failed test case should identify a regression fault, and not just a design change in the frontend.
Manually investigating these false positives can be quite time consuming and expensive.
Therefore, lot of research has been carried out on how to design (and automatically create) more ``robust'' locators.
Some techniques to handle this issue are for example involving multiple locators with a weighted similarity score~\cite{P003}, or by leveraging historical information~\cite{P008}.

The use of locators is essential when writing test cases for web frontend GUIs.
However, as these locators can be quite complex, they can make the test case harder to read and understand.
Furthermore, when a locator needs to be updated due to a design change in the frontend (e.g., a button has been moved into a different position in the page), that update has to be done in every test case using it.
To improve readability and reduce maintenance costs, the
\emph{Page Object}\footnote{https://martinfowler.com/bliki/PageObject.html}
design pattern can be used.
Research work has been carried out to design automated techniques to generate such page-objects~\cite{P140}.

\emph{Flakiness} is a major problem in software testing~\cite{parry2021survey}.
Even when the tested application is not changed, re-running a test case might lead to a different result.
This is a major issue for maintenance, as regression tests might start failing even if no new fault was introduced in the tested application.
There can be many sources of flakiness, especially when dealing with non-deterministic events.
Web applications have their own specific sources of flakiness.
For example, after an event is triggered in a web application, a non-deterministic amount of time might pass before the HTML of page is updated before the next operation in the test case can be executed.
Somehow, a test case has to ``sleep/wait'' some (milli)seconds between two distinct actions.
A timeout that works fine might all of a sudden fail if for any reason a browser takes longer to update a page.
Novel techniques have been designed for example to automatically replace ``sleep'' operations with more robust awaiting mechanisms~\cite{P190}.

\begin{result}
	{\bf \rqCn:} Automated test generation is the most common problem studied in the research literature for web testing. But several other topics have been studied as well, including test maintenance, automated oracles and coverage criteria.
\end{result}

\subsection{\rqDn - \rqD}
\label{rq:D}

In this research question, we look at what kind of contributions the different surveyed articles provide.
This is a \emph{qualitative} categorization, aimed at clustering the different articles based on their contribution.
Each category was first defined by one author, and then refined throughout different iterations with all the other authors.

The categories we used to classify such articles are as following.
Note that an article can provide one or more main contribution categories.
As the name of the category alone might lead to ambiguities in interpretation, we also discuss each of them.

\begin{description}
	\item[New Tool or Technique (NToT):] this represents articles that present a novel tool or technique aimed at testing web applications.
	This category does not include ``tool papers'' (i.e., articles discussing an already published work, by presenting technical details), as those are in a different category.
	\item[Algorithm Experiment (AE):] the articles present empirical studies with tools or techniques. These studies are run by the authors of the article, not involving external human subjects (e.g., students and practitioners).
	\item[Human Experiment (HE):] these articles provide empirical studies with human subjects, which are not the authors of these articles. Usually, those involve studying the human aspects of employing testing techniques and tools in practice.
	\item[Position Paper (PP):] these articles provide reflections, introspections and new ideas, without providing a novel tool or technique. In this group we consider as well ``doctoral symposium'' and ``tool demo'' articles.
	\item[Experience Report (ER):] description of the use of an existing tool/technique in industry, typically without a formal, in-detail empirical study. This would be different from an Algorithm Experiment or Human Experiment done with industrial systems, although the line between them could be considered thin and up to interpretation.
	This group includes ``case study'' and ``action research'' studies, when explicitly mentioned in the text.
	\item[Analysis of Open-Source Projects (OSA):] in these articles, authors do not present novel techniques, but rather studies properties of existing open-source projects (e.g., how are E2E test written? which libraries are used?).
	\item[Survey and Interviews (survey):] surveys of human subjects (e.g., practitioners in industry working with web frontend testing), including as well interviews.
\end{description}

Table~\ref{tab:rq3}
shows each combination of category, with number articles and references.
Table~\ref{tab:rq3_2}
shows each category in isolation.

From this data, we can see that, by far, the most common type of contribution is proposing a new tool or technique for web testing.
In the majority of cases, when in an article a new technique is proposed, it is also empirically validated by the authors.
However, there are still several cases in which no empirical study is carried out when proposing a new technique.
In these cases, the work might mention an empirical study, but little to no information is provided.

Comparatively, there is only few articles involving experiments with human subjects.
Experience reports in industry are even less common.
Analyses of existing practices among open-source projects are only a few.
Surveys and interviews of practitioners are very rare.

From this data we can see a clear trend.
It looks like the research work on web testing is still in an initial academic state, where novel techniques are presented to address several different problems in web testing.
However, these techniques are mainly evaluated ``in the lab'', directly by the authors of these novel techniques.
How practitioners would use such novel techniques and tools, and all the related ``human-aspects'' challenges involved, are not commonly studied in the scientific literature.
These are important research questions that, comparatively, have not received enough attention in the last decade.

\begin{result}
	{\bf \rqDn:} Most of the research work on Web Testing still focus on designing novel techniques and tools, which are then evaluated ``in the lab''. Their application in industry, and the study of their use among human subjects, have received significantly less attention.
\end{result}

\begin{table}[t]
\centering
\caption{Contributions and Total papers per contribution} 
\label{tab:rq3_2}
\begin{tabular}{lr}
  \toprule
Contribution & Nr.Papers \\ 
  \midrule
New Tool or Technique & 201 \\ 
  Algorithm Experiment & 181 \\ 
  Human Experiment & 30 \\ 
  Position Paper & 20 \\ 
  Experience Report & 9 \\ 
  Analysis of Open-Source Projects & 7 \\ 
  Survey and Interviews & 4 \\ 
   \bottomrule
\end{tabular}
\end{table}

\subsection{\rqEn - \rqE}
\label{rq:E}

When designed a novel technique for web testing, or when comparing existing techniques, the choice of which web applications to use for the experiments is of paramount importance.
Without sound empirical analyses, it is not possible to properly evaluate the effectiveness (or lack thereof) of a novel technique.
Therefore, it is important to study how empirical analyses have been carried out in the literature.

In our survey, we have categorized four different groups of application types, as explained next.
An empirical study could comprise several applications of different categories.

\begin{description}
	\item[Artificial:] small, artificial applications developed directly by the authors of these studies.
	%
	\item[Open-Source:] web applications available on open-source repositories (e.g., GitHub), which are then downloaded and started on the local machines of the article authors.
	\item[Online:] web applications available on the internet, that anyone can access.
	\item[Industrial:] industrial web applications, usually tested and used as case study as part of industry-academia collaborations. Note: to fit in this category, and not be labeled under ``Online'', a  collaboration should be explicitly  mentioned in the research articles. This is usually easy to determine when industrial partners are co-authors in these articles. A main difference from ``Online'' category is that here the engineers of these web applications are involved in the studies, and can give feedback to the academics.
\end{description}

Table~\ref{tab:rq5} shows, for each application category, how many articles use at least one of these kind of applications.
Here, for these analysis we consider only the \nAEpapers articles involving ``Algorithm Experiment'' (recall Section~\ref{rq:D}).

\begin{table}[t!]
	\caption{Types of SUTs}
	\label{tab:rq5}
	\centering
\begin{tabular}{l rr rrrr}\\ 
\toprule 
Type & N & \% & Min & Median & Mean & Max  \\ 
\midrule 
Artificial & 27 & 14\% & 1 & 1 & 2.5 & 20 \\ 
 Open-Source & 96 & 53\% & 1 & 5 & 5.2 & 30 \\ 
 Online & 56 & 30\% & 1 & 5 & 156.5 & 6000 \\ 
 Industrial & 31 & 17\% & 1 & 1 & 2.1 & 20 \\ 
 \midrule 
Total & 181 & 100\% & 0 & 4 & 51.9 & 6000 \\ 
 \bottomrule 
\end{tabular} 

\end{table}
As empirical studies might involve more than one application type, we also report the sum of the different types, as a total.
For each entry, we report statistics on how many web applications were used, including the minimum (which by construction would always be at least 1), the arithmetic mean, the median and the max number.

From this data, we can see that artificial examples are not often used, and, when used, there are only a few (mean value $2.5$).
However, in \onlyArtificial of these articles  \emph{only} artificial examples  are used in their empirical studies \refOnlyArtificial.
This poses significant external validity threats for these studies.
The largest use of artificial applications can be found in~\maxArtificialRef, where \maxArtificialValue artificial examples were used.
However, those were
``\emph{J2EE web applications built by professional software engineers}''~\maxArtificialRef.
That study involved the testing of user behavior for synchronous web applications, using Petri nets.
Although \maxArtificialValue artificial examples were used, the study also involved 8  open-source projects.
As these 18 web applications were small (up to 4 554 LOCs, for a total combined of 30 186 LOCs),
``\emph{(to) address scalability, we worked with a company local to our university to apply our technique in an industrial setting}''~\maxArtificialRef.
Artificial examples can be a useful addition to an empirical study, as long as their shortcomings are addressed by using other systems as well.
.

The most common type of applications are open-source, where on average $5.2$ applications are used.
This is a relatively small number, highlighting the difficulty of collecting and enabling suitable web applications for experimentation.
This, unfortunately, might have a negative impact on how likely these proposed novel techniques could fare on new applications.
The largest use of open-source projects can be found in~\maxOpenSourceRef, where \maxOpenSourceValue open-source applications were used.
This study also included 4 artificial applications, for a total of 21 web applications used in the study.

When it comes to web applications available online on internet, a third of studies involves them.
The median of number of online applications used as SUTs is 5 in these studies.
However, the average is very high (175), due to few outliers.
In particular, there are \onlineOutliers articles that involve 1000 or more SUTs \refOnlineOutliers.
On the one hand, not having to download and install applications locally (including setting up all related services, such as for example databases), and just rely on knowledge of the URL of the homepage of a web application, enable the possibility of larger studies.
On the other hand, using online SUTs has its own shortcomings.
First, web applications can change at any time, and so it makes harder to enable the \emph{replication} of these studies.
Second, there might ethical concerns when generating test cases toward an online application, as it can (possibly involuntarily) result in a \emph{denial of service} attack if too many test cases are generated.
The largest use of online applications can be found in~\maxOnlineRef, where \maxOnlineValue applications were used.
These large number of web sites was needed for training machine learning models aimed at
``\emph{allowing the test scripts to automatically adapt to these eventual changes of the web pages}''~\maxOnlineRef, i.e., for test maintainability.

As a sign of maturity of the field, we can see that \nrSutIndustrial studies involve industrial applications.
On average, only $1.5$ industrial applications is used in these studies, with median value being 1.
This means that the majority of these studies only uses 1 industrial application, with only a few having more than 1.
Access to an industrial application not only requires a technique or tool that can actually scale and be applicable on real-world applications, but also considerable effort in establishing industry-academia collaborations~\cite{garousi2019characterizing}.
The largest use of industrial applications can be found in~\maxIndustrialRef, where \maxIndustrialValue industrial web applications were used.
In~\maxIndustrialRef, the Cytestion tool was introduced:
``\emph{an approach that systematically explores a web-based GUI and dynamically builds a test suite for detecting faults that cause visible failures}''.
In the empirical study, 4 open-source projects were used, as well as
\maxIndustrialValue industrial ones.
Those ``\emph{projects were React-based applications developed by different teams of a partner software developing company and referred to a web-integrated business platform, each controlling specific tasks aiming at reducing fiscal risks and costs}''.
Those were large systems, from 80 to 246 thousand LOCs.
Evaluations in industry are essential for sound empirical studies, to verify that academic techniques can actually be usable in practice.
However,  to ease future comparisons, and to reduce validity threats regarding considering applications from only a single enterprise, empirical studies should involve open-source projects as well.
The work in~\maxIndustrialRef is a wonderful example of what can be aimed at for what to use in an empirical study in web testing.

\begin{result}
	{\bf \rqEn: } In algorithm/tool comparisons, on average only $5.2$ open-source projects are used.
	Although this could be considered as a small number, we can see a promising, positive trend in using industrial systems for the empirical studies, done in \nrSutIndustrial articles.
\end{result}

\subsection{\rqFn - \rqF}
\label{rq:F}

Out of \totalPapersIncluded papers, only \totalHEpapers papers conducted experiments involving human subjects such as academics ( students or faculty),  or industry practitioners. In the remaining papers, any experiments conducted were carried out by the authors themselves. There are \totalNumberAcademicParticipants academics and \totalNumberPractitionerParticipants industry practitioners that have participated in these studies. However, it is important to note that papers~\cite{P040} and~\cite{P106} do not specify the number of participants in their experiments.
Table ~\ref{tab:rq9} presents the papers that include experiments conducted by human subjects. These papers explore a variety of categories, which are detailed in Section ~\ref{rq:C}.

While the majority of papers focus on experiments with either academics or industry practitioners,~\cite{P042},~\cite{P052},~\cite{ P095},~\cite{P070},~\cite{P133} and~\cite{P156} have included both.

An empirical study presented in~\cite{P095} evaluates the effectiveness of USherlock, an automated usability evaluation technique. The study involved 24 students and 4 expert evaluators, aiming to assess the adaptability of USherlock for users with varying levels of expertise, including those with medium proficiency.

Recheck, the tool proposed in~\cite{P233}, implements differential testing and was evaluated through an experiment conducted by a professional tester. The evaluation involved a comparison with Selenium WebDriver, and the findings demonstrate that Selenium WebDriver tests equipped with Recheck's oracle mechanisms are more effective in detecting bugs.

In~\cite{P007}  the authors have asked 20 Master students with prior knowledge on web testing to record the time it takes to generate test scripts adopting Page Objects for web application. The students were given the freedom of choosing a web application for testing.

The focus in~\cite{P040} is evaluation the effectivenes of TestComplete tool in GUI Testing. The number of participants is not specified. However it is mentioned that they are senior engineers. The participants had no previous experience with TestComplete.

In~\cite{P070}, the authors address the challenge of concurrent activities in web applications, with a particular focus on the synchronous request-response cycle, utilizing a novel Petri net-based model. The study involved three participants—one from academia and two from industry—tasked with designing and executing tests both with and without the support of WAPG, the tool developed by the authors. The results indicate tht the number of faults detected from the tests developed with WAPG is higher than the number of faults detected from the tests developed by the participants.

In~\cite{P056} the authors focus on automatically generating test data for web application. Two testers were ask to perform this task using the approach introduced by the authors. The authors then compared the time it took each tester to perform the task and concluded that their approach significantly  reduces the time it takes to generate test data.

The exploratory case study presented in~\cite{P103} compares the benefits of scripted and scriptless testing approaches, utilizing Testar and Selenium as representative tools for each method. Among the six participants in the study, only two had prior experience with manual testing. The findings indicate that both approaches offer distinct advantages and are complementary in nature.

The framework proposed in~\cite{P089} aims to enhance the resilience of test cases to UI changes, thereby reducing the costs associated with test maintenance. To evaluate the framework, an experiment was conducted involving three software developers who were tasked with creating a set of test cases for a simple web page. The web page was then manually modified to produce three different versions. The objective was to assess whether the same test cases could be applied across all three versions and to evaluate the impact of these changes on the test cases. The results demonstrate that the framework effectively supports the creation of resilient test cases, provided that the changes made to the application do not alter the core functionality.

In~\cite{P225} five students, either undergraduates or recent graduates, participated in the experiments conducted on five SUTs. Three students had Computer Science background and only one of them had testing experience in the industry. However, the results of the experiments indicated that the tester's background does not influence the tool's effectiveness; instead, it is the tester's level of testing experience that proves to be a more important factor.

Article~\cite{P113} focuses on feature extraction and involves five practioners who have some experience with programming. The participants were tasked with manually extracting features as well as utilizing the tool developed by the authors. The authors subsequently recorded the time taken by the participants in both scenarios to calculate the labor costs. It was conculded that using their tool, the time cost on feature extraction is significantly reduced.

The approach presented in~\cite{P114} focuses on generating test cases from inferred models, implemented through the tool MoLeWe. This approach was evaluated in an experiment involving 18 students, each of whom designed a test suite consisting of ten test cases. These test suites were subsequently used as input for MoLeWe to generate additional test cases. The results demonstrate that the tool is cost-effective and that the test cases generated by MoLeWe significantly increased line coverage compared to manually written tests.

In~\cite{P050}, the authors present a tool designed to enhance state exploration and identify unique URLs. The performance of the proposed tool is compared against two alternative methods—random testing and Q-learning-based testing—as well as with three human participants. Among the participants, one had prior experience in testing, while the other two were novices. The results, based on the number of unique states explored by each method and participant, demonstrated that the tool performs comparably to an experienced human tester. Notably, the tool achieves these results with significantly greater speed.

In~\cite{P052}, the authors introduce a predictive model designed to assist testers in diversifying their testing efforts. To evaluate the model, two separate experiments were conducted. In the first experiment, participants were students enrolled in a software engineering course. The students were divided into 13 groups, each consisting of three members, and tasked with performing exploratory testing on the Cdiscount web application. Each group participated in two 10-minute sessions: one without the tool developed by the authors and one with it. The results demonstrated that the tool effectively supported testers in generating more diverse tests.
In the second experiment, the participants were practitioners from CIS Valley, with varying levels of knowledge in software testing. Similar to the first experiment, the participants conducted one session without the tool and another with it. A survey conducted some time after the experiment, indicated that the tool was again found to be beneficial in enhancing exploratory testing efforts

Article~\cite{P009} introduces WebRR, a record-and-replay tool. While the primary objective of the experiment did not involve human interaction with the tool for conducting tests, the authors engaged a tester from their industrial collaborator to participate. The tester's role was to replicate the tests using both WebRR and Selenium IDE, and record the number of broken tests identified by each tool.

The study in~\cite{P294} focuses specifically on test maintenance and test comprehension. The proposed solution is implemented as a Scout plugin and evaluated through an experiment conducted in collaboration with an industrial partner. Twelve participants with diverse backgrounds were asked to complete a survey assessing and comparing the reports generated with and without the proposed approach.

Scout tool was used also in~\cite{P026}, where the authors investigate the impact of gamification techniques on manual exploratory testing. The study involved 144 participants, all of whom were students enrolled in a Software Engineering course. Most participants had between one and three years of experience in web application development and software testing. Two web applications were selected as SUTs, and the participants conducted tests using the Scout tool, both with and without gamification features enabled. The findings indicate that gamification significantly improves test coverage. However, it was also observed that while test effectiveness remained unaffected, efficiency was negatively impacted by the inclusion of gamification elements.

Three additional studies,~\cite{P042},~\cite{P314} and~\cite{P269} investigate gamification by conducting similar types of experiments involving participants from academia.
In both~\cite{P042} and~\cite{P314} the authors use Scout tool.
In~\cite{P042},   10 participants are recruited, two of which young software developers, whereas the other eight were master students. The participants were divided in two groups, one for each SUTs. Each group conducted the experiment with and without the gamification plugin on Scout tool for 30 minutes.
In~\cite{P314}, the authors investigate the combination of gamification and crowdsourcing for web testing. For the experimental evaluation, six students were recruited and divided into two groups. Each group conducted testing on two websites, one using the proposed approach and the other without it, to assess its effectiveness.

In~\cite{P269} the authors introduce another tool for gamification, GIPGUT. Unlike the previous studies~\cite{P042} and~\cite{P314} where the authors used Scout, in this article the authors build a plugin for IntelliJ IDE. The authors recruited four students to test Amazon website using the proposed tool. The experiment was then followed by a survey to evaluate the performance of the tool and provide feedback.

In~\cite{P039} the authors aim to improve the testability of Web applications  by using hook-based locators. 6 students, organized in three groups,  participated in the experiment. They were responsible to develop a web application with several releases, using the process defined by the authors. Next, the generated test cases with hook-locators were executed and their fragility compared  to the generated test cases with default Katalon locators. It was concluded that hook-based locators reduce the fragility of test cases and of the broken tests.

Article~\cite{P106} focuses on Pattern-based GUI testing. The participants are students, however their exact number is not specified.  The students were assigned to two teams, and each team was assigned one subject study. The students were introduced to PGBT and the PARADIGM-ME tool. Using the PARADIGM-ME tool, the teams began building models to achieve the testing objectives. During the study, they documented the time required to build and configure the models. Next, they generated test cases and executed the tests within the PARADIGM-ME environment. Finally, they analyzed the test results, recorded the identified failures, and concluded the case study.

Article~\cite{P099}  addresses test case prioritization. For the experiment, the authors selected two open-source web applications. 28 students, organized in seven groups were engaged in the experiment. The students were tasked with simulating a realistic software development process using the test automation tool Katalon Studio to create and modify test cases while adapting the application under test (AUT) across multiple iterations. Their responsibilities included generating automated test cases, executing tests, and analyzing the results to build a coverage graph.

The study presented in~\cite{P133} focuses on A/B usability testing and involves the execution of three distinct case studies. In the first case study, two e-commerce web applications were selected as experimental objects, with a participant group of 49 students. The second case study examined a traffic ticketing web application, while the third case study investigated a shoe shopping e-commerce website. For both the second and third case studies, 16 participants were recruited for each experiment. However, the authors do not specify whether these participants were students, practitioners, or individuals from other backgrounds.

Article~\cite{P193} is an opinion survey concentrated  on Selenium WebDriver and its challenges. The participants in this survey were 78 industry professionals experienced in Selenium. The survey results provide valuable insights into the challenges developers and testers encounter, and the strategies they employ to address and resolve issues arising from the use of Selenium. Another survey focused on Selenium is discussed in~\cite{P209}. In this survey a total of 72 industry specialists participated.

In~\cite{P214}, the authors conducted a survey to evaluate the perception and the willingness of practitioners to adopt their proposed approach - a tool designed to assist the development process with Selenium. A total of 148 practitioners participated in the survey, offering valuable insights into the use and perception of WebDriverManager. Another study focused on WebDriverManager is presented in~\cite{P234}, in the context of multi-browser test suites. This study involved 25 master's students and demonstrated that WebDriverManager significantly reduces the time required to set up or update multi-browser test suites.

The study presented in~\cite{P198} conducts an empirical investigation to evaluate the costs and benefits of incorporating Page Object (PO) patterns in web testing. A total of three experiments were carried out, involving 36 participants comprising 19 master's and 17 PhD students. Each participant was tasked with designing a set of test cases both with and without the application of the PO pattern. Following the experiments, the participants completed a questionnaire to provide feedback on their experience. The results indicate that adopting the PO pattern is particularly beneficial in scenarios involving large test suites.

In~\cite{P041}, the authors invited 20 industry practitioners to utilize the framework they developed, MAJD, and subsequently completed a questionnaire designed to assess the ease of use of the framework.

In~\cite{P241}, the authors aim to evaluate the performance of the TESTAR tool within the context of an industrial application developed by SOFTEAM, a French software company. Two SOFTEAM employees participated in the study—a senior analyst and a software developer—both with less than one year of experience in software testing. The participants received training on how to use TESTAR with their application. At the end of the experiment, they completed a questionnaire assessing the tool in terms of quality, learnability, and the likelihood of recommending it.

The authors in~\cite{P265} present VETL, a tool based on a large vision-language model (LVLM) designed for automated test generation. They conducted an experiment to evaluate the tool’s effectiveness in web GUI exploration. Following the experiment, five graduate students with prior experience in GUI testing were engaged. Their role was to assess whether the input text generated by the tool was contextually relevant to the corresponding webpage screenshots.

In~\cite{P276}, the authors propose a novel approach to test generation based on the evaluation of node importance. They formulate two research questions, and to address one of them, they involve five professionals. The participants are asked to assess and assign importance scores to individual nodes, which are then compared to the scores generated by the proposed tool.

The survey in~\cite{P289}investigates developers’ experiences with UI testing in CI/CD pipelines, identifying key challenges, adopted strategies, and perceived impacts on software delivery. To recruit participants, the authors conducted a search on GitHub and contacted 3,338 users via email, inviting them to complete a survey. A total of 128 responses were received, of which 94 were deemed valid for analysis. In addition, the authors conducted follow-up interviews with 18 participants to gain deeper insights.

\begin{result}
	{\bf \rqFn:} Only a limited number of articles, specifically 28, incorporate human studies involving participants from both academia and industry. These participants engage in one or more different stages of the conducted experiments, including performing manual tasks, interacting with testing tools, and providing feedback.
\end{result}

\subsection{\rqGn - \rqG}
\label{rq:G}

A key objective of this survey was to identify the tools introduced over the past decade within the domain of web testing. Out of  \totalPapersIncluded papers we analyzed in this survey, \totalPapersWithTool papers focus on a tool, either newly introduced or improved. Table~\ref{tab:rq6} provides a summary of the tools identified in this study, presenting the tool name (if specified), their open-source status, the number of papers referencing each tool, and a list of those papers.

As discussed in Section~\ref{rq:C}, we categorized papers based on the main challenge they focus on. Hence the tools summarized in Table~\ref{tab:rq6} serve different purposes, e.g., generating test cases, prioritizing test cases, repairing existing test cases.

A total of \totalPapersUnnamedTool  papers classify their associated tools as unnamed. Among these, 10 tools are identified as open-source, while the remaining are not publicly available.

In 3 papers~\cite{P007,P018,P056}, the authors highlight the fact that they are either planning or on the process of implementing the tool.
In~\cite{P007} the authors plan to implement their approach of generating Selenium test suites, as a Selenium plugin. In~\cite{P018}, the challenge the author is tackling is the quality of the front-end  by collecting several metrics. The author mentions that the prototype is being implemented as a client-side Java application. The focus of~\cite{P056} is Test Data Generation for Web Applications. The authors have defined as one of their future work goals to built a tool to automatically generate test data.

It is interesting to note that only 12 tools have more than one paper dedicated. The rest have only one paper published. Here we do not take in consideration the papers that mention or use the tool as a comparison, but only the papers that are dedicated to the tool.
This means we are considering articles in which a tool was first presented, and then all of the following articles in which its authors have extended the tool with new features/techniques or used in new studies.

TESTAR is an automated test generation tool that was originally meant for desktop applications, and later was adapted for web applications too. Therefore the earlier papers on TESTAR,~\cite{P245, P243},  evaluate the tool on both desktop and web applications. A comprehensive overview of TESTAR is provided in~\cite{P108}. The tool has continuoulsy been maintained and improved, as documented in~\cite{P102,P159,P243,P245,P252}. Two industrial case studies are presented in~\cite{P241} and~\cite{P126}. Other papers related to TESTAR, like~\cite{P247} have been excluded from our dataset, as they are not focused on web applications.

APOGEN is a tool that automatically generates Page Objects from web applications. The authors published one journal article~\cite{P140} and two conference papers,~\cite{P200} and~\cite{P208}, all in the same year, 2016.

Model-based regression testing stands at the core of FSMWeb and are discussed in~\cite{P203,P202,P240}.
The tool is primarily focused on regression testing of fail-safe behavior in web application and implements a Genetic Algorithm.

\begin{result}
	{\bf \rqGn:} The large majority of tools are no longer actively maintained after their initial implementation.
	A notable exception is TESTAR.
\end{result}

\subsection{\rqHn - \rqH}
\label{rq:H}

In this section we wanted to investigate the state of open-sourceness of the tools identified and summarized in Table ~\ref{tab:rq6}. We deem this feature to be quite important in the research domain. By making code and methodologies publicly accessible, researchers and developers can validate findings, share innovations, and build upon existing work without barriers.

We have identified \openSourceTools open-source tools. In the papers where tools are marked as open-source, the authors provide a link to the tool (usually a GitHub repository), and/or the experiment results (if any). That means only \ratioYesOpenSource\% of the tools are open-source. In the remaining \notOpenSourceTools papers, the authors do not disclose the tool's implementation, but only their high-level algorithm.

\begin{result}
	{\bf \rqHn:} The majority of the tools are not publicly available, with only \ratioYesOpenSource\% being open-source.
\end{result}

\subsection{\rqIn - \rqI}
\label{rq:I}

Out of the \numberPapersAE papers in which the authors conduct an algorithm experiment, \numberPapersWithComparedTools papers include a comparison with an existing approach. Table ~\ref{tab:rq8} summarizes the tools used for comparison. Notably, Crawljax is the most frequently used benchmark tool, appearing in 10 different studies. Tools, including WebExplor, FraGen, Morpheus, Comjaxtest, SubWeb, Webmole, QExplore, Dante, Awet, and V-DOM, have all been evaluated in  comparison to Crawljax. The prevalence of Crawljax in comparative studies can be attributed, at least in part, to its early introduction in 2008, which has provided it with a longstanding presence in the field.The second most commonly used tool for comparison is WebExplor. Both Crawljax and WebExplor are primarily aimed at automated test generation. WATER, Vista and WebEvo follow as the next most frequently used tools, with a focus on test maintenance.
Additionally, it is noteworthy that seven studies have employed manual testing as a benchmark for evaluating their tools. The diversity of categories, explained in Section ~\ref{rq:D},  for these tools contributes to the limited number of direct comparisons among them.

\begin{result}
	{\bf \rqIn:} Crawljax is the most frequently used  web testing tool for comparison.
\end{result}

\subsection{\rqJn - \rqJ}
\label{rq:J}
As outlined in Section~\ref{rq:D}, our analysis focused on identifying algorithm experiments, human studies, and any experiences authors had in evaluating their proposed approaches or tools. The evaluation metrics varied depending on the paper's objectives and the specific technique or tool under discussion. We paid particular attention to the reported number of faults, as this provided valuable insights into the effectiveness of the proposed solutions. There are  \papersWithFaults articles that discuss their experiments and use number of faults as a metric for evaluation. However \papersNotSpecifiedNrFaults articles do not report the number.
Our findings are summarized in Table~\ref{tab:rq15}. As observed in the table, the majority of studies that conduct experiments, and report the number of identified faults, primarily focus on test generation.

The variation in the number of reported faults is considerable, with~\cite{P145} reporting 3 faults, and~\cite{P005} reporting 3478 faults.

First we will take a look at the papers that report a large number of faults in an attempt to better understand what contributes to that number.

In~\cite{P005}, the authors introduced WebExplor, a tool designed for automated end-to-end web testing. They evaluated the tool on 57 diverse applications, including both real-world and open-source web applications. Conducting an experiment on such a large scale across numerous applications significantly contributed to the high number of reported faults.

VETL, introduced in~\cite{P265}, is a large vision-language model-driven tool designed for end-to-end web testing. It was evaluated on two types of applications: four open-source websites and ten online websites. The number of detected faults varied across applications, with the highest average reaching 1782.4 failures. However, the authors note that the majority of these errors were caused by insufficient memory usage. The second highest number of failures reported for a WUT was 291.2.

TOM is a model-based test generation tool introduced in~\cite{P178}. To evaluate their approach, the authors conducted experiments on a single web application, OntoWorks. Their analysis reported 935 step failures, a considerable number for a single application. However, the authors noted that some behaviors initially classified as failures were, in fact, expected system behavior. Additionally, they highlighted the presence of cascading errors, where a single failure triggered a chain of subsequent failures. While the exact number of actual failures was not specified after these clarifications, these factors likely contributed to the high reported error count.

The study in~\cite{P071} primarily focuses on test maintenance. The authors report a total of 1065 individual test breakages across 453 versions of eight different web applications.

In~\cite{P077}, the tool InwertGen was evaluated on seven web applications, identifying a total of 569 malformed HTML issues. In~\cite{P043}, the proposed framework was tested on 100 web applications, with the authors reporting 255 detected defects.
In~\cite{P002}, the authors report detecting 13 faults in their experiment with an industrial project and 129 faults in open-source projects. In~\cite{P249} the authors report the number of unique bugs, 128, in both mobile and web apps. 

While the majority of papers provide the total number of detected failures, some studies instead report a percentage or an average count.
For instance, in~\cite{P038} ,~\cite{P251} and ~\cite{P145} the authors report the average number of unique failures detected.
In~\cite{P072} and~\cite{P276}, the authors present the fault-finding capability of their implementation as an average percentage.
In~\cite{P032}, the authors emphasize that the reported numbers specifically refer to exception messages rather than other types of failures.
~\cite{P054} further categorizes the detected faults in vulnerabilities and crashes.

\begin{result}
	{\bf RQ10:} The majority of studies reporting the number of detected faults primarily focus on test generation. The reported fault count is affected by factors such as the number of systems under test (SUTs) and the nature of the identified faults.
\end{result}


\subsection{\rqKn - \rqK}
\label{rq:K}

Out of the \totalPapersIncluded papers included in the review, \countWhiteBoxTesting focus on \emph{white-box} testing, \countBlackBoxTesting on \emph{black-box} testing, and the remaining cover not relevant testing techniques for ensuring the quality of web applications. 

The studies focusing on white-box testing techniques extensively cover end-to-end, regression, and unit test automation in web applications~\cite{P015, P044, P188}. The following summary highlights research emphasizing these techniques.

End-to-end and regression testing methods are key aspects explored, with the study in~\cite{P015} emphasizing UI tests and code coverage for transcompiled applications. Similarly, the study in~\cite{P054} generates test cases from client-side behavior models to address server-side vulnerabilities in PHP applications, while~\cite{P083} and~\cite{P167} focus on regression testing automation for PHP applications using GUI and Selenium IDE, respectively.

The studies focusing on unit testing explored coverage metrics and efficiency in test generation for web applications written in various languages. For example, the study in~\cite{P044} targets function, statement, and branch coverage for TypeScript applications, while~\cite{P175} and~\cite{P086} propose advanced techniques for JavaScript applications. 

Similarly, data flow testing is applied to Perl applications in~\cite{P088}, and  full-stack testing approaches, such as in~\cite{P188} and~\cite{P192}, integrate client-server models to classify errors and generate parallel test cases for PHP applications.

Studies such as in~\cite{P147} and~\cite{P221} focus on achieving high coverage in JavaScript testing and assessing adequacy in Web UI testing frameworks for applications in diverse languages, respectively, while specialized techniques like in~\cite{P239} improve client-side testing reliability.

Additionally, studies like in~\cite{P116} emphasize on addressing security vulnerabilities in client-server setups; ~\cite{P256} on domain-specific data use for GUI testing; ~\cite{P261} related to code coverage measurement in industrial web applications; and~\cite{P318} focusing on the use of evolutionary approaches addressing the gap in current black-box tools.
Table~\ref{tab:rq11} provides a detailed description of the studies emphasizing white-box testing.\\

\begin{result}
	{\bf \rqKn:} Our survey reveals a predominant focus on black-box testing, with a notable subset of \countWhiteBoxTesting papers dedicated to white-box testing, particularly emphasizing automation in end-to-end, regression, and unit tests for web applications.
\end{result}


\subsection{\rqLn - \rqL}
\label{rq:L}

We performed a full-text search on the surveyed papers to explore if the popular libraries for GUI testing are mentioned. These libraries - Selenium, Cypress, Puppeteer, and Playwright (recall Section~\ref{sec:background:webtesting}) were used as search keywords.

Among the \totalPapersIncluded reviewed papers, Selenium is mentioned in \selenium papers, Cypress in \cypress, Puppeteer in \puppeteer, and Playwright in \playwright papers.
To examine how the libraries were referenced in the papers, a manual review was conducted on those containing the keywords.
The review revealed that Selenium, Cypress, and Puppeteer were used in fewer papers than mentioned, while Playwright was only cited in the related work sections and not used in any of the studies.  We therefore updated our dataset accordingly and regenerated the corresponding table to reflect these corrections.
Table~\ref{tab:rq12} presents a summary of studies categorized by the types of libraries used for testing.

\begin{result}
	{\bf \rqLn}: Our survey of current studies reveals that Selenium is the most widely used library for enabling GUI testing for web applications.
\end{result}

\subsection{\rqMn - \rqM}
\label{rq:M}

Of the \totalPapersIncluded papers reviewed, \countWithAI employed either search-based or AI-based software testing (SBST/AI) approaches. The automation of web testing has progressed from executing manually written test cases to also generating test cases automatically. Moreover, substantial research efforts have been made not only towards test generation, but also maintenance of tests, test case prioritization, etc, as described in Section~\ref{rq:C}. Considering the complexity of web applications, traditional testing methods face challenges in efficiently and effectively handling dynamic content and frequent updates.

These limitations have led to exploring AI-based solutions.
Table~\ref{tab:rq13} presents a summary of studies categorized by various SBST/AI-based testing methods.

\begin{table}[ht]
\centering
\caption{Existing frameworks and studies utilizing them} 
\label{tab:rq12}
\begin{tabular}{p{4cm}p{2cm}p{7cm}}
  \toprule
Category & Nr.Papers & List.papers \\ 
  \midrule
Selenium & 118 & \cite{P007} \cite{P008} \cite{P011} \cite{P012} \cite{P013} \cite{P016} \cite{P020} \cite{P024} \cite{P029} \cite{P030} \cite{P034} \cite{P037} \cite{P041} \cite{P043} \cite{P047} \cite{P048} \cite{P051} \cite{P053} \cite{P054} \cite{P057} \cite{P064} \cite{P065} \cite{P067} \cite{P069} \cite{P071} \cite{P072} \cite{P075} \cite{P077} \cite{P082} \cite{P084} \cite{P087} \cite{P089} \cite{P090} \cite{P093} \cite{P103} \cite{P104} \cite{P105} \cite{P106} \cite{P108} \cite{P112} \cite{P114} \cite{P119} \cite{P120} \cite{P131} \cite{P134} \cite{P140} \cite{P143} \cite{P144} \cite{P145} \cite{P151} \cite{P155} \cite{P156} \cite{P157} \cite{P164} \cite{P167} \cite{P168} \cite{P169} \cite{P170} \cite{P174} \cite{P178} \cite{P179} \cite{P181} \cite{P182} \cite{P184} \cite{P186} \cite{P189} \cite{P190} \cite{P193} \cite{P194} \cite{P195} \cite{P196} \cite{P197} \cite{P198} \cite{P199} \cite{P200} \cite{P201} \cite{P205} \cite{P206} \cite{P208} \cite{P209} \cite{P210} \cite{P213} \cite{P214} \cite{P215} \cite{P216} \cite{P217} \cite{P218} \cite{P219} \cite{P220} \cite{P221} \cite{P222} \cite{P223} \cite{P225} \cite{P226} \cite{P227} \cite{P228} \cite{P229} \cite{P231} \cite{P232} \cite{P233} \cite{P234} \cite{P235} \cite{P237} \cite{P242} \cite{P254} \cite{P256} \cite{P261} \cite{P265} \cite{P269} \cite{P272} \cite{P283} \cite{P295} \cite{P308} \cite{P312} \cite{P313} \cite{P314} \cite{P315} \cite{P318} \\ 
  Cypress & 5 & \cite{P002} \cite{P031} \cite{P099} \cite{P224} \cite{P280} \\ 
  Puppeteer & 1 & \cite{P038} \\ 
   \bottomrule
\end{tabular}
\end{table}

\begin{table}[t]
\centering
\caption{SBST/AI testing methods and studies utilizing them} 
\label{tab:rq13}
\begin{tabular}{p{4cm}>{\raggedleft}p{2cm}p{7cm}}
  \toprule
Category & \# Papers & List papers \\ 
  \midrule
Search-based &  21 & \cite{P054} \cite{P059} \cite{P067} \cite{P106} \cite{P113} \cite{P116} \cite{P123} \cite{P137} \cite{P143} \cite{P144} \cite{P145} \cite{P192} \cite{P199} \cite{P202} \cite{P203} \cite{P240} \cite{P243} \cite{P266} \cite{P270} \cite{P276} \cite{P318} \\ 
  Reinforcement learning &  14 & \cite{P005} \cite{P029} \cite{P032} \cite{P033} \cite{P050} \cite{P099} \cite{P108} \cite{P126} \cite{P157} \cite{P159} \cite{P241} \cite{P245} \cite{P249} \cite{P251} \\ 
  Computer vision &  13 & \cite{P008} \cite{P013} \cite{P027} \cite{P047} \cite{P048} \cite{P061} \cite{P104} \cite{P154} \cite{P195} \cite{P211} \cite{P219} \cite{P281} \cite{P317} \\ 
  Other ML Models &  11 & \cite{P011} \cite{P062} \cite{P120} \cite{P141} \cite{P173} \cite{P209} \cite{P217} \cite{P226} \cite{P227} \cite{P253} \cite{P284} \\ 
  Generative AI &  11 & \cite{P030} \cite{P112} \cite{P265} \cite{P267} \cite{P280} \cite{P285} \cite{P292} \cite{P293} \cite{P294} \cite{P296} \cite{P308} \\ 
  Natural language processing &  10 & \cite{P004} \cite{P006} \cite{P010} \cite{P052} \cite{P053} \cite{P156} \cite{P197} \cite{P201} \cite{P204} \cite{P215} \\ 
  Reinforcement learning, Computer vision &   1 & \cite{P257} \\ 
   \bottomrule
\end{tabular}
\end{table}

As can be seen in Table~\ref{tab:rq13}, approximately one-quarter of the studies using SBST/AI-based methods employed search-based approaches, while the majority implemented various AI techniques, as outlined below.

\begin{figure}[t!]
	\centering
	\includegraphics[width=0.6\textwidth]{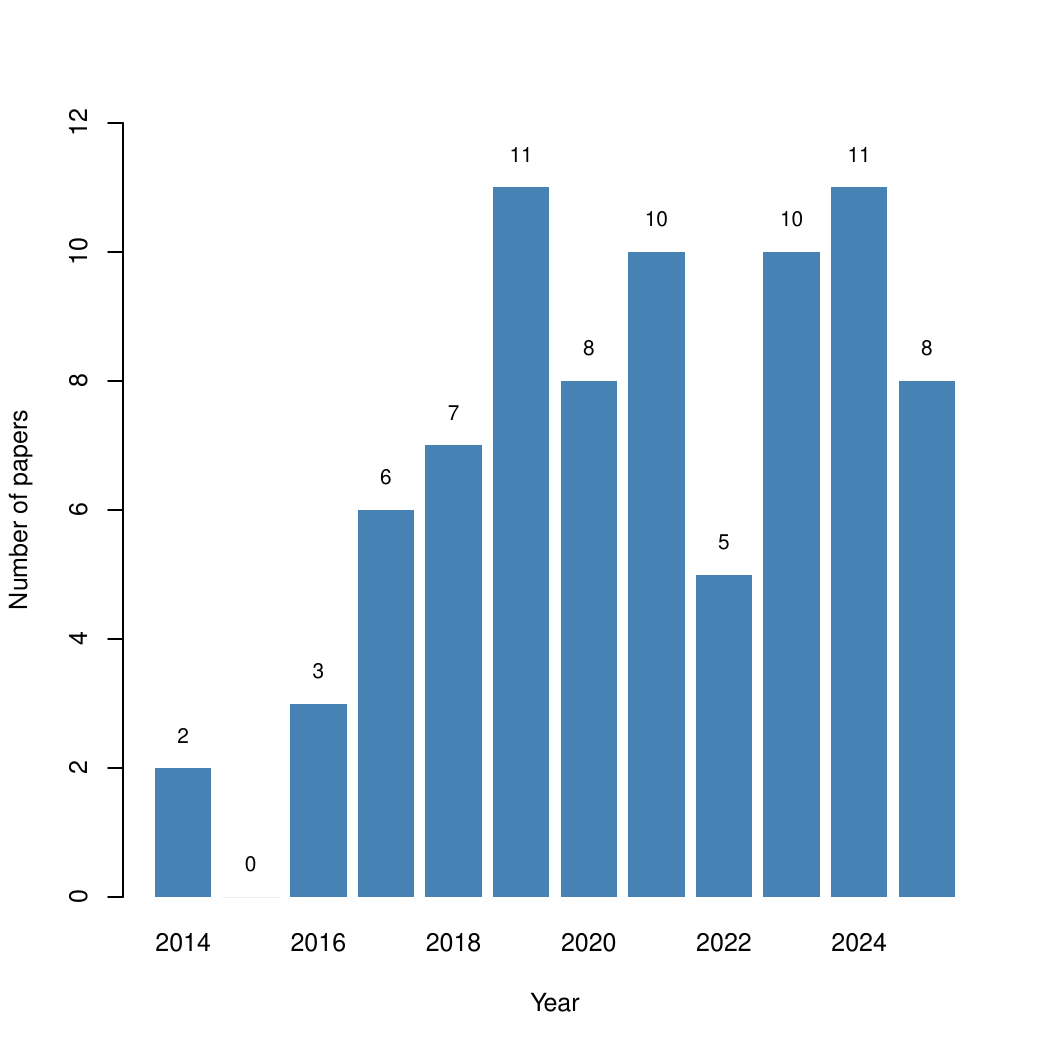}
	\caption{Number of published papers related to the use of SBST/AI-based methods per year} 
	\label{fig:fig_rq13_fig}
\end{figure}
Furthermore, Figure~\ref{fig:fig_rq13_fig}  illustrates the publication rate of papers on the use of SBST/AI-based methods in web testing from January 2014 to May 2025.

\subsubsection{Search-based approaches}

Search-based approaches, particularly Genetic Algorithms, are extensively applied in web application testing for test case generation and UI testing. Studies focus on evolutionary algorithms to enhance client-server testing~\cite{P054}, DOM element crawling~\cite{P067}, and feature extraction~\cite{P113}. Automated test generation integrates Java model simulations with search-based testing~\cite{P137}, while DOM selectors and XPath locators leverage genetic algorithms for UI testing~\cite{P143}. Other studies explore path and input data generation~\cite{P144}, diversity-based generation~\cite{P145}, and parallel evolutionary testing~\cite{P192}. Search-based regression testing and fail-safe behavior testing further expand these techniques~\cite{P202, P203, P240}. Comprehensive end-to-end testing integrates search-based and machine learning solutions~\cite{P199}.

Search-based methods also play a critical role in vulnerability testing, particularly against XML injection attacks. Mutation-based test generation automatically exploits front-end vulnerabilities~\cite{P059}, while COMIX employs genetic algorithms for multi-vulnerability XML injection testing~\cite{P116}. Input data generation for systems under test further ensures robustness in vulnerability detection~\cite{P123}.

Overall, search-based approaches including Genetic Algorithms, are pivotal in web testing, addressing test case generation, UI testing, and regression testing. In parallel, targeted methods efficiently identify vulnerabilities like XML injections, ensuring comprehensive application security~\cite{P054, P067, P123, P199, P059}.

\subsubsection{AI-based approaches}

Regarding AI-based approaches, the studies highlight extensive use of reinforcement learning (particularly Q-learning), computer vision, and natural language processing for testing, in that order. Additionally, various other machine learning approached are also employed using algorithms including neural networks. Details follow.

\begin{itemize}
	\item Reinforcement learning. 
	Reinforcement learning, particularly Q-learning, is extensively applied in web and UI testing, enabling automated test generation~\cite{P005, P032, P108}, prioritization~\cite{P099}, and coverage enhancement~\cite{P157}. Studies explore curiosity-driven exploration~\cite{P005}, visual UI interaction~\cite{P050}, scriptless testing~\cite{P159, P160}, and guided strategies for dynamic applications~\cite{P157, P033}.
	\item Computer vision. 
	Studies utilizing computer vision in web testing focus on UI analysis, automation, and visual segmentation. Approaches include feature mapping~\cite{P013}, widget classification~\cite{P047}, DOM migration via image recognition~\cite{P195}, and deep learning for UI testing~\cite{P061, P104}. Machine learning enhances visual testing~\cite{P027, P154}, with CNNs aiding segmentation and classification~\cite{P219}.
	\item Natural language processing. 
	Studies leveraging natural language processing (NLP) in web application testing focus on automating UI test case generation~\cite{P004}, improving model inference~\cite{P006}, prioritizing test cases~\cite{P010}, and exploratory testing using n-gram models~\cite{P052}. NLP also aids in detecting dependencies~\cite{P197}, identifying web elements~\cite{P201}, repairing test suites~\cite{P204}, and enabling script-free test automation~\cite{P215}.
	\item Computer vision and reinforcement learning combined. Some studies, such as ~\cite{P257}, also combine computer vision and reinforcement learning for automated GUI testing by extracting widgets, analyzing layouts, and representing pages as embedded states. 
\end{itemize}

Additionally, the studies focus on various other machine learning (ML) approaches for web application testing, including UI test automation~\cite{P011}, test generation~\cite{P053, P062}, and performance prediction~\cite{P141}. Methods such as supervised learning~\cite{P011}, deep learning~\cite{P112}, generative adversarial networks (GAN)~\cite{P112}, and support vector machines (SVM)~\cite{P217, P226} are applied for tasks like UI element recognition~\cite{P011}, session-based testing~\cite{P120}, browser environment prioritization~\cite{P062}, and recommendation performance~\cite{P227}.

\begin{result}
	{\bf RQ13:} Our survey of current studies reveals that a significant portion of papers, specifically \countWithAI, used SBST/AI approaches, with about a quarter employing search-based methods and the rest utilizing various AI techniques.
\end{result}

\subsection{\rqNn - \rqN}
\label{rq:N}

As discussed in Section~\ref{rq:D} for \rqDn,
there are \nrER articles that we classified as ``Experience Report''
(recall Table~\ref{tab:rq3_2}), which are \refER.
Furthermore, as discussed in Section~\ref{rq:E} for \rqEn, there are \nrSutIndustrial articles involving algorithm/tool experiments on industrial systems.
The number of those industrial systems in each study ranges from 1 to \maxIndustrialValue (recall Table~\ref{tab:rq5}).
When it comes to studies with human subjects (Section~\ref{rq:F} for \rqFn),
out of \totalHEpapers there are \nrHEIndustrial that involved industrial systems \refHEIndustrial.
The work in~\cite{P108} can be considered a ``tool paper'', where the different features and success stories throughout the years of the GUI (including web) testing tool TESTAR are described.
This includes listing all the different studies in industry where TESTAR was used.
In total, there are \nrIndustrialCollaborators articles in which some form of academia-industry collaboration or industry report was presented, which is \percentageIndustrial\% of the total articles reviewed in this survey.
Table~\ref{tab:collab_papers} lists these \nrIndustrialCollaborators articles.

This data shows a promising trend in this research field, where software engineering results are starting to become usable and able to scale to real-world applications.
This is in contrast with what reported in~\cite{S15}, where it was stated
``\emph{Most papers have little industrial relevance, since industry is not involved in most studies}'', as only
6 articles in their survey  involved web applications from industry.
For an engineering field, the technology transfer from academic research to industrial practice is commendable.
Due to their importance, in the reminder of this section we will discuss these \nrER experience reports in more details.

The work in~\cite{P001} discussed the experience of using
the Automated Testing Framework (ATF) on two clients of PrimeUp,\footnote{https://www.primeup.com.br/}
an IT Services and IT Consulting company in Brazil.
The addressed problem was:
``\emph{Our company faces
	a challenge where each client requires different technologies to
	be employed by our test team to adapt to their needs and applications.
	Thus, our testers must learn different technologies and
	tools for each client}''~\cite{P001}.
To address these challenges, ``\emph{ATF was created by PrimeUp in partnership with Universidade Federal Fluminense
	\ldots which provides a standardized and organized approach for developing
	end-to-end tests for various devices and browsers}''~\cite{P001}.

The work in~\cite{P063} presents the
``\emph{\ldots experience in developing and introducing a set of large
	automated test suites (more than 50 KLOC in total), using best
	practices in state-of-the art and –practice \ldots}''.
The study was done at Innova,\footnote{https://www.innova.com.tr}
a large software firm in Turkey.
It was stated that:
``\emph{Test execution effort (running all test cases) which used to take 2 days was reduced to 1 hour only \ldots
	our industry-academia collaboration and joint work were indeed a success story}''~\cite{P063}.

The article~\cite{P065} was authored by four employees at Varidesk,\footnote{https://www.vari.com/}
a USA-based company that sells office furniture.
That work ``\emph{discusses our efforts at Varidesk to automate web tests against
	our main website}''~\cite{P065}.
One major contribution of that work is that
``\emph{Recognizing that not enough research has been published on what it takes to bring in
	an existing/create a new automation framework, we candidly
	talk about the requirements we needed to meet, and how we
	ultimately went about meeting them}''~\cite{P065}.

WebMate was presented in~\cite{P093}, where the transfer from academic research to a
startup in Germany (Testfabrik\footnote{https://testfabrik.com/en/})
is discussed.
WebMate is a tool aimed at web testing for detecting cross-browser issues.
The article discusses the challenges and lessons learned in creating a commercial product out of academic research in web testing.

The article~\cite{P137} was written by one of us (Prof. Arcuri),
where
``\emph{I report on my direct experience as a PhD/post-doc working in software
	engineering research projects, and then spending the following five years as an engineer in
	two different companies}''~\cite{P137}.
Both these companies are located in Norway (the multi-national
Schlumberger\footnote{https://www.slb.com/}
and
Telenor\footnote{https://www.telenor.no}
).
This reported experience included the discussion of the challenges of manually writing web tests with Selenium and WireMock.
An argument was made that, during that time (2011-2016),
``\emph{research in software engineering, and particularly software testing, has had
	only limited impact on current practices in industry}''~\cite{P137}.

The work in~\cite{P142} presents three case studies in industry, for risk-based testing.
The three involved enterprises are not named in the article.
Some of the involved systems are large web applications with millions of lines of code.

The work in~\cite{P153} discusses
``\emph{ the challenges that we faced in the early
	stages of a technology transfer project aimed to introduce a test
	generator in a software house that develops Web-based Enterprise
	Resource Planning (ERP) solutions}''.
The involved enterprise is not named, and it is described as
``\emph{a mid-size Italian software house}''~\cite{P153}.
Different challenges are discussed in the article, including regarding scalability, lack of oracles and test reporting.

Testinium\footnote{https://testinium.com/} is a large software testing company in Turkey.
In the work in~\cite{P170}, they report on their use of the Gauge\footnote{https://gauge.org/index.html}
framework in several of their clients.
Gauge is an open-source automated testing framework for behavior-driven development for web applications.

As discussed in Section~\ref{rq:C}, flakiness in web test cases is a major source of problems.
In~\cite{P210},
its authors
``\emph{were involved in the refactoring of an existing automated
	flaky E2E test suite for a large Web application. In this paper, we report
	on our experience \ldots Our procedure allowed to reduce the flakiness to virtually zero}''.
The involved organization was PRINTO,\footnote{https://www.printo.it/}
a non-profit international research medical network.\\

\begin{result}
	{\bf \rqNn:} Out of \totalPapersIncluded surveyed articles, there are \nrSutIndustrial algorithm/tool studies using industrial systems, \nrHEIndustrial studies with human subjects on industrial systems, 1 tool paper and \nrER experience reports in industry.
	Although \percentageIndustrial\% still represents a minority of the studies, this shows a promising, positive trend in the research field of web testing.
\end{result}


\section{Discussion}
\label{sec:discussion}

The large number of publications (i.e., \totalPapersIncluded) in the period of 2014-2025 shows the importance of the topic of web testing in the research community (\rqAn).
Many of these articles were published in the leading conferences and journals in software engineering research (\rqBn), like for example FSE and ICSE, and not just specialized venues for software testing like ICST.
This provides evidence on the maturity and quality of research work done on this topic.

The field of web testing covers several different aspects, although test case generation is the most common (\rqCn).
This is an important complex problem that is still not fully solved, with scalable solutions that are widely adopted in industry.
More research is needed.

When doing research, there are different kinds of studies that can be carried out.
In the context of web testing, by far the most common case is to design and present a novel technique, and evaluate in in the lab on a set of web applications (\rqDn).
Other kinds of studies, like for example with human subjects or analyses of existing practices, are comparatively fewer.
This creates a large gap in the literature that needs to be filled.

Most studies in the literature use open-source projects as artifacts for the empirical evaluations (\rqEn).
Although there is a positive trend in using systems from industrial collaborations in the studies, the number of  used applications is typically small.
This creates major threats to the external validity of many studies in the literature.
Unfortunately, setting up and configuring real-world web applications for empirical studies is no trivial task.
There is a need and an opportunity here to create easy-to-reuse collections of web applications to foster and simplify future studies in web testing.

As already mentioned, studies with human subjects are just a few, and several of those only involve students (\rqFn).
This kind of studies are expensive, time-consuming and risky to carry out, especially when done in industry with industrial practitioners, which explains their rarity.
However, they are essential to study how techniques only used in the lab by their authors can actually be employed in the real-world.
To get a better technology transfer from academic results to industrial practice, more of this kind of studies are needed.

The large majority of studies involve designing novel techniques.
These techniques are often implemented in software tools (\rqGn), but only 1/3 of them seems released as open-source (\rqHn).
This is a major issue, as it can hinder comparisons with existing work (\rqIn).
Comparisons with the state-of-the-art is essential to properly evaluate newly proposed techniques.
Furthermore, unless commercial web testing tools integrate academic techniques evaluated in research papers, the impact of these latter on industrial practice would be limited.
Even in the case of open-source tools, these are often used in just a few studies (typically only one), and then discontinued by their academic authors (e.g., PhD students graduating and moving on to different research topics).
Mature academic open-source tools that are developed throughout the years, used in several different studies, and that can be used by practitioners in industry, seems a rarity.
A notable exception is TESTAR.
For a better impact of academic research on industrial practice, there is a need of more success stories like TESTAR.

Regarding impact, even if academic tools are not directly used by practitioners in industry (e.g., due to usability concerns and lack of maintenance), automatically finding and reporting faults in existing systems can provide concrete benefits (\rqJn).
However, test case generation is only one problem in the research field of web testing.
How to increase the industry impact of other kinds of studies is an important problem to address.

Testing a web application from its GUI frontend requires using a browser and interacting with HTML pages.
However, the behavior of the application strongly depend on the software that is run on the frontend (e.g., JavaScript event handlers) and what running in the backend (including databases).
However, very few techniques and studies in the literature exploit the knowledge of the source code when testing a web application (\rqKn), in the so-called white-box testing.
As the analysis of the source code of an application can lead to better testing results (e.g., considering the literature of white-box testing in other domains besides web testing), this creates a major research opportunity for designing novel techniques that can further push forward the state-of-the-art.

To interact with a browser in a test case, specialized libraries are needed.
Those are Selenium, Cypress, Puppeteer, and Playwright.
However, in academic studies, Selenium is by the far the most used, whereas the other libraries are nearly ignored (\rqLn).
Considering the widespread use of these other libraries in industry, this might hinder technology transfer from academic results to industrial practice.
Why academics focus only on Selenium, and whether the choice of a library is just a technical detail and not relevant for research work, is something that would need more investigation.

Testing web applications is an important topic that has been investigated for at least the last three decades.
It is composed of many complex problems, from test generation to test maintenance.
The use of AI techniques might provide working solutions for several of these challenging research problems.
As such, the use of AI techniques has seen a major increase in the academic literature in the last few years (\rqMn).
How to best use these techniques, and how to combine them (e.g., SBST+LMM) to get better results, are important research topics of high potential, especially when applied in contexts that have seen only little work in the literature so far, like white-box testing (\rqKn).

Research in a practical engineering field such as software engineering should, in the long term, be applicable and scalable to real-world applications and contexts.
Direct applications of academic results in industry via academia-industry collaborations can help reaching such goals.
In our survey, we have seen that \percentageIndustrial\% of the reviewed articles involved some sort application in industry (\rqNn).
This shows a positive, inspiring trend.
The use of innovative techniques like AI (\rqMn) might be the key to solve real-world problems that can be of value for practitioners in industry (\rqNn).

\section{Threats To Validity}
\label{sec:threats}

To ensure the validity of our findings, we have taken various measures to minimize potential biases. However, some limitations remain, which we discuss in this section. 
A potential limitation of this study is the paper selection process. We conducted searches in the most widely used academic databases, with search queries carefully reviewed and agreed upon by all authors. Additionally, we performed both forward and backward snowballing to identify further relevant studies. Since this process was manually conducted, we implemented a validation step where at least two authors reviewed the final list of selected papers to minimize errors and reduce the risk of omitting relevant studies. However, despite these efforts, there remains a possibility that some relevant papers on web testing may have been missed. Additionally, there remains a possibility that other researchers may obtain different results in the selection of relevant papers, leading to variations in the final list of included studies.

A second potential limitation of this study is the data extraction process. While data analysis was conducted rigorously, all manually extracted data was verified by at least two authors. In cases of disagreement, additional authors reviewed the paper, and the final extracted data was determined through consensus. Certain qualitative data points, such as category and contribution, were defined by the authors. Although we have taken great care to ensure accuracy in assigning labels and definitions, there remains the possibility of misinterpretation. Additionally, other researchers may have differing perspectives on these classifications.

\section{Conclusions}
\label{sec:conclusions}

In this study, we analyzed \totalPapersIncluded papers published over the past decade (2014-2025), focusing on web testing.
We searched through major databases and reviewed every paper to ensure its relevance.
Our research questions were designed to provide a comprehensive and detailed understanding of the trends and current state of web testing.

The volume of publications indicates an interest from the research community, with annual publication ranging from 10 to 31 papers across various conferences and journals, notably ICST.
These studies cover a broad range of topics and study categories.
While both black-box and white-box testing are utilized in research, black-box testing is significantly more prevalent.  Additionally, we examined trends in the adoption of AI techniques.
Our findings also highlight the development of numerous web testing tools, some of which have been released as open-source solutions.
Furthermore, many of these studies include empirical evaluations, ranging from algorithmic experiments to human participant studies, or a combination of both.
The experiments were primarily conducted in academic settings.
However, there is a growing trend of collaboration with industry.

This survey offers valuable insights into the evolution of web testing, key trends, challenges, and advancements in the field.
Future research can build on these findings, expanding open-source availability of web testing tools, to further refine testing methodologies and explore emerging technologies that enhance the efficiency and effectiveness of web application testing.

\section*{CRediT authorship contribution statement}
\textbf{Iva Kertusha:}  Writing - Original Draft, Writing - Review \& Editing,  Methodology, Investigation, Data Curation, Visualization,
\textbf{Gebremariam Assres:} Writing - Original Draft,Writing - Review \& Editing,
\textbf{Onur Duman:} Investigation, Data curation,
\textbf{Andrea Arcuri:} Writing - Review \& Editing, Supervision, Methodology, Conceptualization

\section*{Declaration of interests}
The authors declare that they have no known competing financial interests or personal relationships that could have appeared to influence the work reported in this paper.

\section*{Acknowledgments}
This work is funded by the European Research Council (ERC) under the European Union’s Horizon 2020 research and innovation programme (EAST project, grant agreement No. 864972).

\clearpage
\appendix
\section{Supplementary tables}
\label{appendix}

\begin{table}[H]
\centering
\caption{Common Publishing Venues} 
\label{tab:rq2}
\begin{tabular}{lrl}
  \toprule
Venue & Nr.Papers & Venue.Type \\ 
  \midrule
Other conferences & 66 & conference \\ 
  ICST & 28 & conference \\ 
  Other journals & 24 & Journal \\ 
  FSE & 11 & conference \\ 
  ICSE & 10 & conference \\ 
  ICSTW & 10 & workshop; conference \\ 
  JSS & 10 & journal \\ 
  STVR & 8 & journal \\ 
  TSE & 7 & journal \\ 
  ASE & 6 & conference \\ 
  ISSTA & 6 & conference \\ 
  QUATIC & 6 & conference \\ 
  SQJ & 6 & journal \\ 
  AST & 5 & conference; workshop \\ 
  EMSE & 5 & journal \\ 
  IST & 5 & journal \\ 
  ICSME & 4 & conference \\ 
  TOSEM & 4 & journal \\ 
  COMPSAC & 3 & conference \\ 
  ICWE & 3 & conference \\ 
  \textless Programming\textgreater & 2 & conference \\ 
  AUSE & 2 & journal \\ 
  DSA & 2 & conference \\ 
  ESEM & 2 & conference \\ 
  ICEBE & 2 & conference \\ 
  IEEE Access & 2 & journal \\ 
  IJSEKE & 2 & journal \\ 
  ISMSIT & 2 & conference \\ 
  ISSRE & 2 & conference \\ 
  JSEP & 2 & journal \\ 
  SoSyM & 2 & journal \\ 
  SSBSE & 2 & conference \\ 
   \bottomrule
\end{tabular}
\end{table}

\begin{longtable}{p{4cm}>{\raggedleft\arraybackslash}p{2cm}p{7cm}}
\caption{List of categories and respective papers} \\ 
  \toprule
Category & Nr.Papers & List.papers \\ 
  \midrule
Test Generation & 90 & \cite{P002} \cite{P005} \cite{P007} \cite{P011} \cite{P013} \cite{P017} \cite{P021} \cite{P022} \cite{P023} \cite{P024} \cite{P029} \cite{P030} \cite{P032} \cite{P033} \cite{P034} \cite{P038} \cite{P040} \cite{P047} \cite{P054} \cite{P056} \cite{P059} \cite{P066} \cite{P073} \cite{P074} \cite{P077} \cite{P079} \cite{P080} \cite{P084} \cite{P087} \cite{P088} \cite{P091} \cite{P103} \cite{P106} \cite{P108} \cite{P112} \cite{P116} \cite{P120} \cite{P123} \cite{P126} \cite{P131} \cite{P134} \cite{P137} \cite{P141} \cite{P147} \cite{P153} \cite{P156} \cite{P159} \cite{P167} \cite{P178} \cite{P180} \cite{P181} \cite{P187} \cite{P188} \cite{P192} \cite{P202} \cite{P203} \cite{P217} \cite{P219} \cite{P222} \cite{P225} \cite{P226} \cite{P227} \cite{P228} \cite{P229} \cite{P235} \cite{P239} \cite{P240} \cite{P241} \cite{P243} \cite{P244} \cite{P246} \cite{P248} \cite{P249} \cite{P251} \cite{P252} \cite{P256} \cite{P257} \cite{P265} \cite{P270} \cite{P277} \cite{P280} \cite{P283} \cite{P285} \cite{P295} \cite{P297} \cite{P298} \cite{P302} \cite{P305} \cite{P312} \cite{P318} \\ 
  Test Maintenance & 47 & \cite{P003} \cite{P004} \cite{P008} \cite{P009} \cite{P012} \cite{P039} \cite{P058} \cite{P064} \cite{P082} \cite{P083} \cite{P089} \cite{P090} \cite{P105} \cite{P115} \cite{P119} \cite{P143} \cite{P179} \cite{P182} \cite{P183} \cite{P184} \cite{P185} \cite{P190} \cite{P194} \cite{P195} \cite{P196} \cite{P198} \cite{P199} \cite{P200} \cite{P201} \cite{P204} \cite{P207} \cite{P208} \cite{P210} \cite{P212} \cite{P223} \cite{P232} \cite{P238} \cite{P242} \cite{P250} \cite{P253} \cite{P254} \cite{P267} \cite{P268} \cite{P289} \cite{P292} \cite{P294} \cite{P296} \\ 
  Testing Framework & 44 & \cite{P001} \cite{P016} \cite{P026} \cite{P028} \cite{P043} \cite{P045} \cite{P046} \cite{P049} \cite{P051} \cite{P063} \cite{P065} \cite{P069} \cite{P076} \cite{P078} \cite{P094} \cite{P128} \cite{P133} \cite{P142} \cite{P155} \cite{P164} \cite{P170} \cite{P186} \cite{P193} \cite{P197} \cite{P205} \cite{P206} \cite{P209} \cite{P213} \cite{P214} \cite{P215} \cite{P216} \cite{P218} \cite{P220} \cite{P224} \cite{P231} \cite{P234} \cite{P269} \cite{P272} \cite{P274} \cite{P281} \cite{P293} \cite{P313} \cite{P314} \cite{P315} \\ 
  Test Generation; Model Inference & 18 & \cite{P037} \cite{P050} \cite{P052} \cite{P053} \cite{P068} \cite{P075} \cite{P097} \cite{P113} \cite{P114} \cite{P118} \cite{P136} \cite{P144} \cite{P145} \cite{P157} \cite{P163} \cite{P168} \cite{P173} \cite{P236} \\ 
  Oracle & 10 & \cite{P014} \cite{P048} \cite{P057} \cite{P061} \cite{P072} \cite{P095} \cite{P151} \cite{P154} \cite{P174} \cite{P233} \\ 
  Coverage Criteria & 8 & \cite{P015} \cite{P018} \cite{P019} \cite{P086} \cite{P132} \cite{P189} \cite{P230} \cite{P261} \\ 
  Model Inference & 7 & \cite{P006} \cite{P020} \cite{P025} \cite{P102} \cite{P150} \cite{P284} \cite{P317} \\ 
  Test Prioritization & 5 & \cite{P010} \cite{P060} \cite{P062} \cite{P099} \cite{P278} \\ 
  Test Generation; Model Inference; Oracle & 3 & \cite{P035} \cite{P067} \cite{P169} \\ 
  Test Generation; Coverage Criteria & 3 & \cite{P135} \cite{P175} \cite{P245} \\ 
  Model Inference; Oracle & 2 & \cite{P027} \cite{P092} \\ 
  Dataset & 2 & \cite{P031} \cite{P299} \\ 
  Test Generation; Oracle & 2 & \cite{P093} \cite{P237} \\ 
  Test Generation; Test Maintenance & 2 & \cite{P104} \cite{P211} \\ 
  Testing Framework; Test Generation & 1 & \cite{P041} \\ 
  Testing Framework; Oracle & 1 & \cite{P042} \\ 
  Test Generation; Model Inference; Coverage Criteria & 1 & \cite{P070} \\ 
  Taxonomy; Test Maintenance & 1 & \cite{P071} \\ 
  Taxonomy; Oracle & 1 & \cite{P085} \\ 
  Taxonomy & 1 & \cite{P110} \\ 
  Test Generation; Testing Framework & 1 & \cite{P111} \\ 
  Test Generation; Testing Framework; Model Inference & 1 & \cite{P117} \\ 
  Testing Framework; Oracle & 1 & \cite{P124} \\ 
  Test Maintenance; Model Inference & 1 & \cite{P140} \\ 
  Coverage Criteria; Taxonomy & 1 & \cite{P221} \\ 
  Test Prioritization; Coverage Criteria & 1 & \cite{P266} \\ 
  Model Inference; Oracle & 1 & \cite{P273} \\ 
  Test Generation; Test Prioritization & 1 & \cite{P276} \\ 
  Model Inference; Test Maintenance & 1 & \cite{P287} \\ 
  Test Generation & 1 & \cite{P308} \\ 
   \bottomrule

\label{tab:rq4}
\end{longtable}

\begin{table}[htpb]
\centering
\caption{List of papers for every contribution} 
\label{tab:rq3}
\begin{tabular}{p{2cm}>{\raggedleft\arraybackslash}p{2cm}p{9cm}}
  \toprule
Type & Nr.Papers & List.papers \\ 
  \midrule
NToT;AE & 162 & \cite{P002} \cite{P003} \cite{P004} \cite{P005} \cite{P006} \cite{P008} \cite{P010} \cite{P011} \cite{P013} \cite{P014} \cite{P015} \cite{P016} \cite{P017} \cite{P018} \cite{P021} \cite{P022} \cite{P024} \cite{P025} \cite{P028} \cite{P029} \cite{P030} \cite{P032} \cite{P033} \cite{P035} \cite{P037} \cite{P038} \cite{P043} \cite{P047} \cite{P048} \cite{P053} \cite{P054} \cite{P057} \cite{P058} \cite{P059} \cite{P060} \cite{P061} \cite{P062} \cite{P066} \cite{P067} \cite{P072} \cite{P073} \cite{P075} \cite{P077} \cite{P079} \cite{P080} \cite{P082} \cite{P083} \cite{P085} \cite{P086} \cite{P087} \cite{P088} \cite{P092} \cite{P094} \cite{P097} \cite{P102} \cite{P104} \cite{P112} \cite{P115} \cite{P116} \cite{P117} \cite{P118} \cite{P119} \cite{P120} \cite{P123} \cite{P124} \cite{P126} \cite{P131} \cite{P132} \cite{P135} \cite{P136} \cite{P140} \cite{P141} \cite{P143} \cite{P144} \cite{P145} \cite{P147} \cite{P150} \cite{P151} \cite{P154} \cite{P155} \cite{P157} \cite{P167} \cite{P169} \cite{P173} \cite{P174} \cite{P175} \cite{P178} \cite{P179} \cite{P180} \cite{P181} \cite{P182} \cite{P183} \cite{P184} \cite{P185} \cite{P186} \cite{P187} \cite{P188} \cite{P189} \cite{P190} \cite{P192} \cite{P194} \cite{P195} \cite{P196} \cite{P197} \cite{P200} \cite{P201} \cite{P202} \cite{P203} \cite{P204} \cite{P205} \cite{P206} \cite{P213} \cite{P217} \cite{P218} \cite{P219} \cite{P222} \cite{P223} \cite{P226} \cite{P227} \cite{P228} \cite{P229} \cite{P230} \cite{P231} \cite{P232} \cite{P235} \cite{P236} \cite{P237} \cite{P239} \cite{P240} \cite{P242} \cite{P243} \cite{P244} \cite{P245} \cite{P246} \cite{P248} \cite{P249} \cite{P250} \cite{P251} \cite{P252} \cite{P253} \cite{P254} \cite{P256} \cite{P257} \cite{P266} \cite{P267} \cite{P268} \cite{P270} \cite{P273} \cite{P278} \cite{P280} \cite{P281} \cite{P284} \cite{P292} \cite{P293} \cite{P295} \cite{P296} \cite{P298} \cite{P302} \cite{P305} \cite{P308} \cite{P312} \cite{P315} \\ 
  PP & 20 & \cite{P012} \cite{P020} \cite{P031} \cite{P046} \cite{P108} \cite{P111} \cite{P159} \cite{P163} \cite{P199} \cite{P207} \cite{P208} \cite{P211} \cite{P216} \cite{P224} \cite{P261} \cite{P274} \cite{P287} \cite{P297} \cite{P299} \cite{P318} \\ 
  NToT & 17 & \cite{P019} \cite{P023} \cite{P027} \cite{P034} \cite{P064} \cite{P068} \cite{P069} \cite{P074} \cite{P076} \cite{P084} \cite{P091} \cite{P164} \cite{P168} \cite{P272} \cite{P277} \cite{P283} \cite{P313} \\ 
  NToT;HE & 13 & \cite{P009} \cite{P039} \cite{P041} \cite{P042} \cite{P052} \cite{P056} \cite{P070} \cite{P089} \cite{P106} \cite{P114} \cite{P133} \cite{P269} \cite{P314} \\ 
  HE & 10 & \cite{P007} \cite{P026} \cite{P040} \cite{P095} \cite{P103} \cite{P156} \cite{P198} \cite{P233} \cite{P234} \cite{P241} \\ 
  AE & 10 & \cite{P045} \cite{P049} \cite{P051} \cite{P078} \cite{P105} \cite{P128} \cite{P215} \cite{P220} \cite{P285} \cite{P317} \\ 
  ER & 9 & \cite{P001} \cite{P063} \cite{P065} \cite{P093} \cite{P137} \cite{P142} \cite{P153} \cite{P170} \cite{P210} \\ 
  NToT;AE;HE & 7 & \cite{P050} \cite{P099} \cite{P113} \cite{P225} \cite{P265} \cite{P276} \cite{P294} \\ 
  OSA & 5 & \cite{P071} \cite{P090} \cite{P110} \cite{P212} \cite{P238} \\ 
  survey & 3 & \cite{P193} \cite{P209} \cite{P289} \\ 
  AE;OSA & 1 & \cite{P134} \\ 
  NToT;survey & 1 & \cite{P214} \\ 
  NToT;AE;OSA & 1 & \cite{P221} \\ 
   \bottomrule
\end{tabular}
\end{table}

\clearpage
\begin{longtable}{p{3cm}p{1cm}>{\raggedleft}p{3cm}>{\raggedleft}p{3cm}r}
\caption{Number of participants} \\ 
  \toprule
Category & Ref & \# academic & \# practitioners & \# total \\ 
  \midrule
Oracle & \cite{P095} & 24 & 2 & 26 \\ 
  Oracle & \cite{P233} & 0 & 1 & 1 \\ 
  Test Generation & \cite{P007} & 20 & 0 & 20 \\ 
  Test Generation & \cite{P040} & 0 & not specified & not specified \\ 
  Test Generation & \cite{P056} & 0 & 4 & 4 \\ 
  Test Generation & \cite{P103} & 0 & 6 & 6 \\ 
  Test Generation & \cite{P106} & not specified & 0 & not specified \\ 
  Test Generation & \cite{P156} & 1 & 3 & 4 \\ 
  Test Generation & \cite{P225} & 5 & 0 & 5 \\ 
  Test Generation & \cite{P241} & 0 & 2 & 2 \\ 
  Test Generation & \cite{P265} & 5 & 0 & 5 \\ 
  Test Generation; Model Inference & \cite{P050} & 0 & 3 & 3 \\ 
  Test Generation; Model Inference & \cite{P052} & 39 & 8 & 47 \\ 
  Test Generation; Model Inference & \cite{P113} & 0 & 5 & 5 \\ 
  Test Generation; Model Inference & \cite{P114} & 18 & 0 & 18 \\ 
  Test Generation; Model Inference; Coverage Criteria & \cite{P070} & 1 & 2 & 3 \\ 
  Test Generation; Test Prioritization & \cite{P276} & 0 & 5 & 5 \\ 
  Test Maintenance & \cite{P009} & 0 & 1 & 1 \\ 
  Test Maintenance & \cite{P039} & 6 & 0 & 6 \\ 
  Test Maintenance & \cite{P089} & 0 & 3 & 3 \\ 
  Test Maintenance & \cite{P198} & 36 & 0 & 36 \\ 
  Test Maintenance & \cite{P294} & 0 & 12 & 12 \\ 
  Test Maintenance & \cite{P289} & 0 & 94 & 94 \\ 
  Test Prioritization & \cite{P099} & 28 & 0 & 28 \\ 
  Testing Framework & \cite{P026} & 144 & 0 & 144 \\ 
  Testing Framework & \cite{P133} & 49 & 32 & 81 \\ 
  Testing Framework & \cite{P234} & 25 & 0 & 25 \\ 
  Testing Framework & \cite{P269} & 4 & 0 & 4 \\ 
  Testing Framework & \cite{P314} & 6 & 0 & 6 \\ 
  Testing Framework & \cite{P193} & 0 & 78 & 78 \\ 
  Testing Framework & \cite{P209} & 0 & 72 & 72 \\ 
  Testing Framework & \cite{P214} & 0 & 148 & 148 \\ 
  Testing Framework; Oracle & \cite{P042} & 8 & 2 & 10 \\ 
  Testing Framework; Test Generation & \cite{P041} & 0 & 20 & 20 \\ 

\bottomrule
\label{tab:rq9}
\end{longtable}

\clearpage
\begin{longtable}{p{4cm}p{3cm}>{\raggedleft}p{2cm}p{4cm}}
\caption{Introduced or Improved tools} \\ 
  \toprule
Tool & Open Source & \# Papers & List papers \\ 
  \midrule
UNNAMED & No &  61 & \cite{P006} \cite{P011} \cite{P013} \cite{P026} \cite{P027} \cite{P029} \cite{P039} \cite{P042} \cite{P043} \cite{P047} \cite{P048} \cite{P050} \cite{P052} \cite{P053} \cite{P054} \cite{P059} \cite{P060} \cite{P061} \cite{P065} \cite{P066} \cite{P069} \cite{P076} \cite{P080} \cite{P083} \cite{P084} \cite{P087} \cite{P088} \cite{P089} \cite{P094} \cite{P104} \cite{P112} \cite{P113} \cite{P115} \cite{P119} \cite{P120} \cite{P123} \cite{P131} \cite{P143} \cite{P154} \cite{P164} \cite{P167} \cite{P168} \cite{P169} \cite{P173} \cite{P186} \cite{P189} \cite{P192} \cite{P217} \cite{P218} \cite{P219} \cite{P222} \cite{P226} \cite{P227} \cite{P228} \cite{P229} \cite{P250} \cite{P253} \cite{P254} \cite{P266} \cite{P276} \cite{P296} \\ 
  UNNAMED & Yes &  10 & \cite{P014} \cite{P028} \cite{P030} \cite{P062} \cite{P067} \cite{P184} \cite{P201} \cite{P281} \cite{P292} \cite{P312} \\ 
  TESTAR & Yes &   8 & \cite{P102} \cite{P108} \cite{P126} \cite{P159} \cite{P241} \cite{P243} \cite{P245} \cite{P252} \\ 
  Apogen & Yes &   4 & \cite{P140} \cite{P200} \cite{P208} \cite{P242} \\ 
  FSMWeb & No &   3 & \cite{P202} \cite{P203} \cite{P240} \\ 
  Not implemented Yet & NA &   3 & \cite{P007} \cite{P018} \cite{P056} \\ 
  Cornpickle & Yes &   2 & \cite{P085} \cite{P246} \\ 
  E2E-LOADER & Yes &   2 & \cite{P298} \cite{P305} \\ 
  PESTO & Yes &   2 & \cite{P155} \cite{P195} \\ 
  PIRLTest & Yes &   2 & \cite{P249} \cite{P257} \\ 
  ViMoTest & Yes &   2 & \cite{P297} \cite{P302} \\ 
  Vista & Yes &   2 & \cite{P012} \cite{P182} \\ 
  WAPG & No &   2 & \cite{P070} \cite{P135} \\ 
  X-BROT & No &   2 & \cite{P057} \cite{P174} \\ 
  AJAS & No &   1 & \cite{P068} \\ 
  ApiCarv & Yes &   1 & \cite{P024} \\ 
  Assessor & Yes &   1 & \cite{P232} \\ 
  ATA & No &   1 & \cite{P082} \\ 
  ATRINA & Yes &   1 & \cite{P072} \\ 
  Automation Xtreme & No &   1 & \cite{P313} \\ 
  autoOrchTest & No &   1 & \cite{P074} \\ 
  AWET & Yes &   1 & \cite{P225} \\ 
  COLOR & No &   1 & \cite{P058} \\ 
  COMIX & No &   1 & \cite{P116} \\ 
  ComJaxTest & No &   1 & \cite{P118} \\ 
  Confix & No &   1 & \cite{P017} \\ 
  CrawLabel & Yes &   1 & \cite{P004} \\ 
  Crawljax & No &   1 & \cite{P073} \\ 
  Cypress Copilot & Yes &   1 & \cite{P280} \\ 
  Cytestion & Yes &   1 & \cite{P002} \\ 
  DANTE & Yes &   1 & \cite{P187} \\ 
  DBInputs & Yes &   1 & \cite{P256} \\ 
  DIG & Yes &   1 & \cite{P145} \\ 
  E2EGIT &  &   1 & \cite{P299} \\ 
  eBat & Yes &   1 & \cite{P038} \\ 
  EventBreak & No &   1 & \cite{P021} \\ 
  EvoMaster & Yes &   1 & \cite{P318} \\ 
  FOREPOST & No &   1 & \cite{P141} \\ 
  FraGen & Yes &   1 & \cite{P035} \\ 
  Fraunhofer Tool & No &   1 & \cite{P075} \\ 
  FRET & No &   1 & \cite{P315} \\ 
  FsCheck & No &   1 & \cite{P136} \\ 
  FuzzE & Yes &   1 & \cite{P295} \\ 
  GIPGUT & Yes &   1 & \cite{P269} \\ 
  GTpql & Yes &   1 & \cite{P033} \\ 
  GUITAR & Yes &   1 & \cite{P163} \\ 
  history-diagnostics & Yes &   1 & \cite{P023} \\ 
  InwertGen & No &   1 & \cite{P077} \\ 
  jFAT & No &   1 & \cite{P231} \\ 
  JSdep & Yes &   1 & \cite{P180} \\ 
  JSEFT & Yes &   1 & \cite{P175} \\ 
  Judge & Yes &   1 & \cite{P284} \\ 
  KeyDriver & No &   1 & \cite{P091} \\ 
  KeyjaxTest & No &   1 & \cite{P097} \\ 
  Kraken 2.0 & Yes &   1 & \cite{P111} \\ 
  Link & No &   1 & \cite{P022} \\ 
  LOTUS & No &   1 & \cite{P237} \\ 
  MAEWU & Yes &   1 & \cite{P221} \\ 
  MAJD & Yes &   1 & \cite{P041} \\ 
  MARG & No &   1 & \cite{P251} \\ 
  MBTCover & Yes &   1 & \cite{P261} \\ 
  MBUITC & Yes &   1 & \cite{P117} \\ 
  MoLeWe & Yes &   1 & \cite{P114} \\ 
  Morpheus Web Testing & Yes &   1 & \cite{P037} \\ 
  Mutandis & Yes &   1 & \cite{P086} \\ 
  OOGSSA & No &   1 & \cite{P270} \\ 
  P-GUI & No &   1 & \cite{P016} \\ 
  PARADIGM-ME & No &   1 & \cite{P106} \\ 
  PARTE & No &   1 & \cite{P235} \\ 
  PathFinder & Yes &   1 & \cite{P293} \\ 
  Pattern-Based Usability Testing & No &   1 & \cite{P124} \\ 
  QExplore & Yes &   1 & \cite{P157} \\ 
  RClassify & No &   1 & \cite{P239} \\ 
  ReDeCheck & No &   1 & \cite{P151} \\ 
  RLTCP & No &   1 & \cite{P099} \\ 
  Robot ( framework) & No &   1 & \cite{P034} \\ 
  ROBULA+ & Yes &   1 & \cite{P179} \\ 
  Scout plugin & No &   1 & \cite{P314} \\ 
  Scout-LLM & Yes &   1 & \cite{P294} \\ 
  ScPlay, ScRec, ScEdit & No &   1 & \cite{P133} \\ 
  Selenium-Jupiter & Yes &   1 & \cite{P213} \\ 
  SEMTER & No &   1 & \cite{P204} \\ 
  SgaTest & No &   1 & \cite{P273} \\ 
  Sibilla & No &   1 & \cite{P025} \\ 
  SideeX & No &   1 & \cite{P064} \\ 
  SIDEREAL & No &   1 & \cite{P196} \\ 
  Similo & Yes &   1 & \cite{P003} \\ 
  SkyFire & Yes &   1 & \cite{P244} \\ 
  SleepReplacer & Yes &   1 & \cite{P190} \\ 
  StackFul & No &   1 & \cite{P188} \\ 
  Stile & Yes &   1 & \cite{P205} \\ 
  Subweb & No &   1 & \cite{P144} \\ 
  SymJS & No &   1 & \cite{P147} \\ 
  Tansuo & No &   1 & \cite{P236} \\ 
  TEDD & Yes &   1 & \cite{P197} \\ 
  TERMINATOR & No &   1 & \cite{P010} \\ 
  TesMa & No &   1 & \cite{P079} \\ 
  TEST-Web & Yes &   1 & \cite{P272} \\ 
  Testilizer & Yes &   1 & \cite{P181} \\ 
  TOM & Yes &   1 & \cite{P178} \\ 
  TransCompiledCodeCoverage & Yes &   1 & \cite{P015} \\ 
  UAES & No &   1 & \cite{P268} \\ 
  UITESTFIX & No &   1 & \cite{P183} \\ 
  V-DOM & No &   1 & \cite{P230} \\ 
  VETL & Yes &   1 & \cite{P265} \\ 
  VON Similo LLM & Yes &   1 & \cite{P267} \\ 
  Waterfall & No &   1 & \cite{P194} \\ 
  WebDriverManager & Yes &   1 & \cite{P214} \\ 
  WebEV & Yes &   1 & \cite{P031} \\ 
  WebEvo & Yes &   1 & \cite{P008} \\ 
  WebExplor & No &   1 & \cite{P005} \\ 
  WebMate & No &   1 & \cite{P093} \\ 
  WEBMOLE & Yes &   1 & \cite{P150} \\ 
  WebQT & No &   1 & \cite{P032} \\ 
  WebRR & No &   1 & \cite{P009} \\ 
  WebTest & Yes &   1 & \cite{P132} \\ 
  WebTestRepair & No &   1 & \cite{P223} \\ 
  WebTestSuiteRepair & No &   1 & \cite{P185} \\ 
\bottomrule
\label{tab:rq6}
\end{longtable}

\clearpage
\begin{longtable}{p{4cm}>{\raggedleft}p{2cm}p{6cm}}
\caption{List of tools used for comparison} \\ 
  \toprule
Tool & \# Papers & List papers \\ 
  \midrule
Crawljax & 10 & \cite{P005} \cite{P035} \cite{P037} \cite{P118} \cite{P144} \cite{P150} \cite{P157} \cite{P187} \cite{P225} \cite{P230} \\ 
  WebExplor & 8 & \cite{P005} \cite{P032} \cite{P038} \cite{P157} \cite{P249} \cite{P251} \cite{P257} \cite{P265} \\ 
  WATER & 8 & \cite{P008} \cite{P012} \cite{P182} \cite{P183} \cite{P184} \cite{P194} \cite{P204} \cite{P292} \\ 
  Vista & 4 & \cite{P008} \cite{P009} \cite{P183} \cite{P204} \\ 
  Selenium & 3 & \cite{P009} \cite{P115} \cite{P215} \\ 
  WebEvo & 3 & \cite{P183} \cite{P204} \cite{P292} \\ 
  Subweb & 2 & \cite{P005} \cite{P145} \\ 
  Robula+ & 2 & \cite{P115} \cite{P196} \\ 
  Atusa & 2 & \cite{P145} \cite{P187} \\ 
  Artemis & 2 & \cite{P147} \cite{P180} \\ 
  SFTM & 2 & \cite{P183} \cite{P292} \\ 
  TESTAR & 1 & \cite{P002} \\ 
  LML & 1 & \cite{P003} \\ 
  DIG & 1 & \cite{P005} \\ 
  Random & 1 & \cite{P005} \\ 
  WATERFALL & 1 & \cite{P008} \\ 
  Atlassian Clover & 1 & \cite{P015} \\ 
  EclEmma & 1 & \cite{P015} \\ 
  monkey testing & 1 & \cite{P030} \\ 
  GPT-3.5-Turbo & 1 & \cite{P030} \\ 
  WebQTse & 1 & \cite{P032} \\ 
  WebQTr & 1 & \cite{P032} \\ 
  MUTANDIS & 1 & \cite{P072} \\ 
  FeedEx & 1 & \cite{P097} \\ 
  FireCrystal & 1 & \cite{P113} \\ 
  FireCrow & 1 & \cite{P113} \\ 
  vMOSA & 1 & \cite{P116} \\ 
  MBUITC & 1 & \cite{P117} \\ 
  VeriWeb & 1 & \cite{P118} \\ 
  Web 1.0 Crawler & 1 & \cite{P150} \\ 
  Depth-First Crawler & 1 & \cite{P150} \\ 
  Tansuo & 1 & \cite{P150} \\ 
  WebExplorer & 1 & \cite{P157} \\ 
  ARTEMIS & 1 & \cite{P175} \\ 
  Montoto & 1 & \cite{P196} \\ 
  Selenium WebDriver & 1 & \cite{P215} \\ 
  Selenium IDE & 1 & \cite{P215} \\ 
  unnamed commercial & 1 & \cite{P215} \\ 
  APOGEN & 1 & \cite{P222} \\ 
  DANTE & 1 & \cite{P225} \\ 
  WebSphinx & 1 & \cite{P236} \\ 
  Link Checker Pro & 1 & \cite{P236} \\ 
  Axe DevTools Pro & 1 & \cite{P237} \\ 
  EVENTRACER & 1 & \cite{P239} \\ 
  R4 & 1 & \cite{P239} \\ 
  Monkey & 1 & \cite{P249} \\ 
  QExplore & 1 & \cite{P251} \\ 
  Katalon & 1 & \cite{P254} \\ 
  LINK & 1 & \cite{P256} \\ 
  KeyjaxTest & 1 & \cite{P273} \\ 
  YOLO & 1 & \cite{P281} \\ 
  WebEmbed & 1 & \cite{P284} \\ 
  GraphMAE & 1 & \cite{P284} \\ 
  ContentHash & 1 & \cite{P284} \\ 
  Levenshtein & 1 & \cite{P284} \\ 
  RTED & 1 & \cite{P284} \\ 
  SimHash & 1 & \cite{P284} \\ 
  BlockHash & 1 & \cite{P284} \\ 
  Pdiff & 1 & \cite{P284} \\ 
  Phash & 1 & \cite{P284} \\ 
  Histogram & 1 & \cite{P284} \\ 
  SIFT & 1 & \cite{P284} \\ 
  SSIM & 1 & \cite{P284} \\ 
  FragGen & 1 & \cite{P284} \\ 
  VISTA & 1 & \cite{P292} \\ 
  EDIT DIS & 1 & \cite{P292} \\ 
  SleepReplacer & 1 & \cite{P296} \\ 
\bottomrule
\label{tab:rq8}
\end{longtable}

\clearpage
\begin{longtable}{p{1cm}p{5cm}p{3cm}r}
\caption{Report number of faults} \\ 
  \toprule
Ref & Title & Category & \# Reported Faults \\ 
  \midrule
\cite{P145} & Diversity-Based Web Test Generation & Test Generation; Model Inference & 3 \\ 
  \cite{P126} & Automated Testing of Web Applications with TESTAR: Lessons Learned Testing the Odoo Tool & Test Generation & 5 \\ 
  \cite{P021} & EventBreak: analyzing the responsiveness of user interfaces through performance-guided test generation & Test Generation & 6 \\ 
  \cite{P134} & Test them all, is it worth it? Assessing configuration sampling on the JHipster Web development stack & Test Generation & 6 \\ 
  \cite{P163} & GUITAR: an innovative tool for automated testing of GUI-driven software & Test Generation; Model Inference & 7 \\ 
  \cite{P136} & Property-based testing of web services by deriving properties from business-rule models & Test Generation; Model Inference & 8 \\ 
  \cite{P241} & Evaluating the TESTAR tool in an Industrial Case Study & Test Generation & 11 \\ 
  \cite{P103} & Scripted and scriptless GUI testing for web applications: An industrial case & Test Generation & 12 \\ 
  \cite{P245} & Q-learning strategies for action selection in the TESTAR
automated testing tool & Test Generation; Coverage Criteria & 12 \\ 
  \cite{P243} & Using genetic programming to evolve action selection rules in traversal-based automated software testing: results obtained with the TESTAR tool & Test Generation & 13 \\ 
  \cite{P257} & Effective, Platform-Independent GUI Testing via Image Embedding and Reinforcement Learning & Test Generation & 13 \\ 
  \cite{P038} & eBAT: An Efficient Automated Web Application Testing Approach Based on Tester’s Behavior & Test Generation & 17.8 \\ 
  \cite{P118} & Automated Testing of Web Applications Using Combinatorial Strategies & Test Generation; Model Inference & 18 \\ 
  \cite{P060} & A Method for Test Cases Reduction in Web Application Testing Based on User Session & Test Prioritization & 20 \\ 
  \cite{P169} & Input Contract Testing of Graphical User Interfaces & Test Generation; Model Inference; Oracle & 25 \\ 
  \cite{P054} & Test Case Generation Based on Client-Server of Web Applications by Memetic Algorithm & Test Generation & 26 \\ 
  \cite{P188} & Prioritising Server Side Reachability via Inter-process Concolic Testing & Test Generation & 26 \\ 
  \cite{P175} & JSEFT: Automated JavaScript Unit Test Generation & Test Generation; Coverage Criteria & 35 \\ 
  \cite{P049} & Behavior-driven development (BDD) Cucumber Katalon for Automation GUI testing case CURA and Swag Labs & Testing Framework & 36 \\ 
  \cite{P070} & Using Petri Nets to Test Concurrent Behavior of Web Applications & Test Generation; Model Inference; Coverage Criteria & 36 \\ 
  \cite{P228} & Website Functionality Testing Using the Automation Tool & Test Generation & 36 \\ 
  \cite{P251} & Can Cooperative Multi-Agent Reinforcement Learning Boost Automatic Web Testing? An Exploratory Study & Test Generation & 39.5 \\ 
  \cite{P295} & FuzzE, Development of a Fuzzing Approach for Odoo's Tours Integration Testing Plateform & Test Generation & 45 \\ 
  \cite{P135} & Testing concurrent user behavior of synchronous web applications with Petri nets & Test Generation; Coverage Criteria & 57 \\ 
  \cite{P032} & A Reinforcement Learning Approach to Generating Test Cases for Web Applications & Test Generation & 69 \\ 
  \cite{P157} & QExplore: An exploration strategy for dynamic web applications using guided search & Test Generation; Model Inference & 69 \\ 
  \cite{P072} & Atrina: Inferring Unit Oracles from GUI Test Cases & Oracle & 76 \\ 
  \cite{P276} & Multi-Objective Test Case Generation for Web Applications with Limited Resources & Test Generation; Test Prioritization & 80 \\ 
  \cite{P315} & Tree-Based Synthesis of Web Test Sequences from Manual Actions & Testing Framework & 86 \\ 
  \cite{P085} & Testing Web Applications Through Layout Constraints & Taxonomy; Oracle & 90 \\ 
  \cite{P094} & Event-driven web application testing based on model-based mutation testing & Testing Framework & 92 \\ 
  \cite{P249} & Effective, Platform-Independent GUI Testing via Image Embedding and Reinforcement Learning & Test Generation & 128 \\ 
  \cite{P002} & Cytestion: Automated GUI Testing for Web Applications & Test Generation & 142 \\ 
  \cite{P106} & Pattern-based GUI testing: Bridging the gap between design and quality assurance & Test Generation & 148 \\ 
  \cite{P095} & Empirical validation of an automatic usability evaluation method & Oracle & 195 \\ 
  \cite{P043} & A Selenium-based Web Application Automation Test Framework & Testing Framework & 255 \\ 
  \cite{P077} & Incremental Web Application Testing Using Page Object & Test Generation & 569 \\ 
  \cite{P066} & Combinatorial Testing of Full Text Search in Web Applications & Test Generation & 797 \\ 
  \cite{P178} & TOM: a Model-Based GUI Testing framework & Test Generation & 935 \\ 
  \cite{P071} & Why do Record/Replay Tests of Web Applications Break? & Taxonomy; Test Maintenance & 1065 \\ 
  \cite{P265} & Leveraging Large Vision Language Model For Better Automatic Web GUI Testing & Test Generation & 1780 \\ 
  \cite{P005} & Automatic Web Testing Using Curiosity-Driven Reinforcement Learning & Test Generation & 3478 \\ 
  \cite{P040} & Evaluation for Web GUI Automation Testing Tool - Experiment & Test Generation & not specified \\ 
  \cite{P042} & Gamified Exploratory GUI Testing of Web Applications: a Preliminary Evaluation & Testing Framework; Oracle & not specified \\ 
  \cite{P045} & Test Driven Development in Action: Case Study of a Cross-Platform Web Application & Testing Framework & not specified \\ 
  \cite{P050} & Automating GUI Testing with Image-Based Deep Reinforcement Learning & Test Generation; Model Inference & not specified \\ 
  \cite{P059} & Automatic Generation of Tests to Exploit XML Injection Vulnerabilities in Web Applications & Test Generation & not specified \\ 
  \cite{P075} & Model generation to support model-based testing applied on the NASA DAT Web-application - An experience report & Test Generation; Model Inference & not specified \\ 
  \cite{P076} & A framework for testing web applications using action word based testing & Testing Framework & not specified \\ 
  \cite{P080} & Syntax-based test case generation for web application & Test Generation & not specified \\ 
  \cite{P091} & A Keyword-Driven Tool for Testing Web Applications (KeyDriver) & Test Generation & not specified \\ 
  \cite{P093} & WebMate: Web Application Test Generation in the Real World & Test Generation; Oracle & not specified \\ 
  \cite{P104} & Using deep learning for selenium web UI functional tests: A case-study with e-commerce applications & Test Generation; Test Maintenance & not specified \\ 
  \cite{P108} & testar – scriptless testing through graphical user interface & Test Generation & not specified \\ 
  \cite{P113} & Test-Driven Feature Extraction of Web Components & Test Generation; Model Inference & not specified \\ 
  \cite{P114} & Model-based testing leveraged for automated web tests & Test Generation; Model Inference & not specified \\ 
  \cite{P117} & Automated Model-Based Test Case Generation for Web User Interfaces (WUI) From Interaction Flow Modeling Language (IFML) Models & Test Generation; Testing Framework; Model Inference & not specified \\ 
  \cite{P132} & Exploring output-based coverage for testing PHP web applications & Coverage Criteria & not specified \\ 
  \cite{P154} & Artificial intelligence in automated system for web-interfaces visual testing & Oracle & not specified \\ 
  \cite{P229} & A Semantic Web Enabled Approach to Automate Test Script Generation for Web Applications & Test Generation & not specified \\ 
  \cite{P244} & Skyfire: Model-Based Testing with Cucumber & Test Generation & not specified \\ 
  \cite{P246} & Testing Web Applications Through Layout Constraints
 & Test Generation & not specified \\ 
  \cite{P266} & Segment-Based Test Case Prioritization: A Multi-objective Approach & Test Prioritization; Coverage Criteria & not specified \\ 
  \cite{P270} & A Hybrid Approach for Automated GUI Testing
using Quasi-Oppositional Genetic Sparrow Search
Algorithm & Test Generation & not specified \\ 
  \cite{P272} & TEST-Web: An Automated Web Application
Testing Tool for Enhanced Efficiency and
Consistency & Testing Framework & not specified \\ 
  \cite{P278} & A Regression Test Case Prioritization Technique for Web Application Using User Session Data & Test Prioritization & not specified \\ 
  \cite{P297} & ViMoTest: A Tool to Specify ViewModel-Based GUI
Test Scenarios using Projectional Editing & Test Generation & not specified \\ 
  \cite{P314} & Conceptualization of Multi-user Collaborative GUI-Testing for Web Applications & Testing Framework & not specified \\ 
   \bottomrule
\label{tab:rq15}
\end{longtable}

\clearpage
\begin{longtable}{p{1cm}p{3cm}p{9cm}}
\caption{RQ11 - List of studies using white-box testing techniques with highlights} \\ 
  \toprule
Citation & Title & Study.highlights \\ 
  \midrule
\cite{P015} & Code coverage for any kind of test in any kind of transcompiled cross-platform applications & Presents a framework-independent Java tool for calculating code coverage for cross-platform applications. The tool offers versatility but slower performance compared to established unit-testing tools. \\ 
  \cite{P054} & Test Case Generation Based on Client-Server of Web Applications by Memetic Algorithm & Proposes a search-based testing approach using memetic algorithm to link client-side models with server-side code, enhancing vulnerability detection in web applications. \\ 
  \cite{P072} & Atrina: Inferring Unit Oracles from GUI Test Cases & Introduces a test tool, ATRINA, that leverages DOM-based assertions and execution data to generate accurate unit-level assertions for JavaScript. The tool outperforms human-written and mutation-based methods in fault detection. \\ 
  \cite{P083} & Novel approach to reuse unused test cases in a GUI based application & Addresses GUI regression testing by reusing unused test paths with varied inputs, improving test feasibility, reducing costs, and saving time for modified applications with frequent user-driven changes. \\ 
  \cite{P086} & Guided Mutation Testing for JavaScript Web Applications & Introduces a mutation testing tool, MUTANDIS, for JavaScript. The tool optimizes mutation generation using static and dynamic analysis, and ranking functions by importance, significantly reducing computational costs and improving mutant relevance. \\ 
  \cite{P088} & Data Flow Testing of CGI Based Web Applications & Proposes a data flow testing technique for CGI programs in Perl, modeling definition-use chains, identifying test paths for variable interactions, and enhancing the quality of web applications handling user input and database interfaces. \\ 
  \cite{P116} & Search-based multi-vulnerability testing of XML injections in web applications & Presents a search-based approach, COMIX, for uncovering multiple web service vulnerabilities simultaneously. This approach improves security testing for interactions between web applications and back-end services. \\ 
  \cite{P147} & SymJS: Automatic Symbolic Testing of JavaScript Web Applications & Introduces a framework (called SymJS) for automated client-side JavaScript testing that combines symbolic execution, dynamic feedback, and taint analysis to achieve high test coverage. \\ 
  \cite{P167} & Automatically Generating Test Scripts for GUI Testing
 & Proposes a method leveraging static and dynamic analysis of application source code and executables to automate test script generation. The method achieves a substantial reduction in man-hours compared to traditional manual script creation. \\ 
  \cite{P175} & JSEFT: Automated JavaScript Unit Test Generation & Presents a framework (called JSEFT) for generating test cases for JavaScript applications at both event and function levels. It outperforms existing test automation frameworks in achieving high coverage and fault detection accuracy. \\ 
  \cite{P188} & Prioritising Server Side Reachability via Inter-process Concolic Testing & Introduces a concolic tester called StackFul for full-stack JavaScript web applications. This tool enables to distinguish high- and low-priority server errors by applying a whole-program perspective thereby improving error classification and testing efficiency. \\ 
  \cite{P189} & Comparing coverage criteria for dynamic web application: An empirical evaluation & Compares the effectiveness of dynamic test coverage criteria in web applications through mutation analysis. Results show that DOM coverage is the most effective and efficient, followed by virtual DOM, HTML element, and statement coverage criteria. \\ 
  \cite{P192} & Parallel evolutionary test case generation for web applications & Introduces an approach based on parallel genetic algorithm for web application test case generation. This approach improves- efficiency by reducing execution time and iterations, and effectiveness through population diversity using an island model and migration mechanism. \\ 
  \cite{P221} & Mutation analysis for assessing end-to-end web tests & Presents a mutation analysis framework (called MAEWU) that evaluates web UI test suites by mutating dynamic DOM elements. Results show its effectiveness in assessing test suite adequacy and improving web app test quality. \\ 
  \cite{P235} & An efficient regression testing approach for PHP web applications: a controlled experiment & Proposes a regression testing approach using impact analysis and program slicing to efficiently test modified code areas in web applications. Results show cost reduction in regression testing, depending on application characteristics and update frequency. \\ 
  \cite{P236} & Using combinatorial testing to build navigation graphs for dynamic web applications & Presents a combinatorial approach (called Tansuo) for building navigation graphs in dynamic web applications. The approach generates test sequences, achieving code coverage outperforming other navigation graph tools in effectiveness. \\ 
  \cite{P239} & RClassify: classifying race conditions in web applications via deterministic replay & Presents a deterministic replay-based method for identifying harmful race conditions in web applications. The method significantly reducing false positives and outperforming existing techniques in detecting real, impactful races. \\ 
  \cite{P249} & Effective, Platform-Independent GUI Testing via Image Embedding and Reinforcement Learning & This study focuses on code coverage and instrumentation and suggests vision-based and reinforcement learning approaches, including platform-independent frameworks to improve the effectiveness, efficiency, and generalizability of automated GUI testing across mobile and web applications. \\ 
  \cite{P256} & 
DBInputs: Exploiting Persistent Data to Improve Automated GUI Testing & The authors suggest that exploiting domain-specific data from application databases and improving test generation efficiency through techniques like input reuse and Q-learning can enhance automated GUI testing. \\ 
  \cite{P261} & Coverage measurement in model-based testing of web
applications: Tool support and an industrial experience report  & This study proposes an automated model-based testing (MBT) tool, called MBTCover. The tool enables live measurement of requirements and model coverage, as well as code coverage, in large scale industrial web application testing projects. \\ 
  \cite{P313} & Automation Xtreme - A Web Automation AI Tool & This study encompasses both black-box and white-box testing and introduces an AI-driven web automation tool called Xtreme-A. The tool aims at optimizing web testing through intelligent automation, enhancing efficiency, and minimizing manual effort. \\ 
  \cite{P318} & Search-Based White-Box Fuzzing of Web Frontend Applications & In this study, the author emphasizes the importance of frontend testing for improved user experience and proposes a novel white-box approach using evolutionary algorithms. Existing tools are mostly black-box, limiting fault detection and no white-box tools currently exist.
 \\ 
\bottomrule
\label{tab:rq11}
\end{longtable}

\clearpage
\begin{longtable}{p{9cm}p{1cm}p{3cm}}
\caption{Papers with Industrial Collaborations} \\ 
  \toprule
Title & Ref & Category \\ 
  \midrule
ATF - An end-to-end testing framework: experience report & \cite{P001} & Testing Framework \\ 
  Cytestion: Automated GUI Testing for Web Applications & \cite{P002} & Test Generation \\ 
  Automatic Web Testing Using Curiosity-Driven Reinforcement Learning & \cite{P005} & Test Generation \\ 
  WebRR: self-replay enhanced robust record/replay for web application testing & \cite{P009} & Test Maintenance \\ 
  TERMINATOR: better automated UI test case prioritization & \cite{P010} & Test Prioritization \\ 
  Beyond Page Objects: Testing Web Applications with State Objects: Use states to drive your tests & \cite{P020} & Model Inference \\ 
  Morpheus Web Testing: A Tool for Generating Test Cases for Widget Based Web Applications & \cite{P037} & Test Generation; Model Inference \\ 
  Image-based Approaches for Automating GUI Testing of Interactive Web-based Applications & \cite{P048} & Oracle \\ 
  Web Application Testing With Model Based Testing Method: Case Study & \cite{P051} & Testing Framework \\ 
  Fostering the Diversity of Exploratory Testing in Web Applications & \cite{P052} & Test Generation; Model Inference \\ 
  Automatic Generation of Tests to Exploit XML Injection Vulnerabilities in Web Applications & \cite{P059} & Test Generation \\ 
  Automating Web Application Testing from the Ground Up: Experiences and Lessons Learned in an Industrial Setting & \cite{P065} & Testing Framework \\ 
  Combinatorial Testing of Full Text Search in Web Applications & \cite{P066} & Test Generation \\ 
  Using Semantic Similarity in Crawling-Based Web Application Testing & \cite{P067} & Test Generation; Model Inference; Oracle \\ 
  Model generation to support model-based testing applied on the NASA DAT Web-application - An experience report & \cite{P075} & Test Generation; Model Inference \\ 
  WebMate: Web Application Test Generation in the Real World & \cite{P093} & Test Generation; Oracle \\ 
  Scripted and scriptless GUI testing for web applications: An industrial case & \cite{P103} & Test Generation \\ 
  testar – scriptless testing through graphical user interface & \cite{P108} & Test Generation \\ 
  Search-based multi-vulnerability testing of XML injections in web applications & \cite{P116} & Test Generation \\ 
  A Search-Based Testing Approach for XML Injection Vulnerabilities in Web Applications & \cite{P123} & Test Generation \\ 
  Testing concurrent user behavior of synchronous web applications with Petri nets & \cite{P135} & Test Generation; Coverage Criteria \\ 
  Property-based testing of web services by deriving properties from business-rule models & \cite{P136} & Test Generation; Model Inference \\ 
  An experience report on applying software testing academic results in industry: we need usable automated test generation & \cite{P137} & Test Generation \\ 
  FOREPOST: finding performance problems automatically with feedback-directed learning software testing & \cite{P141} & Test Generation \\ 
  A multiple case study on risk-based testing in industry
 & \cite{P142} & Testing Framework \\ 
  GUI Testing in Production: Challenges and Opportunities & \cite{P153} & Test Generation \\ 
  QExplore: An exploration strategy for dynamic web applications using guided search & \cite{P157} & Test Generation; Model Inference \\ 
  Input Contract Testing of Graphical User Interfaces & \cite{P169} & Test Generation; Model Inference; Oracle \\ 
  Test Automation with the Gauge Framework: Experience and Best Practices & \cite{P170} & Testing Framework \\ 
  Automated Fixing of Web UI Tests via Iterative Element Matching & \cite{P183} & Test Maintenance \\ 
  An automated model-based approach to repair test suites of evolving
web applications & \cite{P184} & Test Maintenance \\ 
  SleepReplacer: a novel tool‑based approach for replacing thread sleeps in selenium WebDriver test code & \cite{P190} & Test Maintenance \\ 
  Black-box model-based regression testing of fail-safe behavior in web applications & \cite{P202} & Test Generation \\ 
  A case study of black box fail-safe testing in web applications & \cite{P203} & Test Generation \\ 
  Reducing Flakiness in End-to-End Test Suites: An Experience Report & \cite{P210} & Test Maintenance \\ 
  Automating end-to-end web testing via manual testing & \cite{P222} & Test Generation \\ 
  Evaluating the TESTAR tool in an Industrial Case Study & \cite{P241} & Test Generation \\ 
  Skyfire: Model-Based Testing with Cucumber & \cite{P244} & Test Generation \\ 
  Towards a Robust Waiting Strategy for Web GUI Testing for an Industrial Software System & \cite{P250} & Test Maintenance \\ 
  Coverage measurement in model-based testing of web
applications: Tool support and an industrial experience report  & \cite{P261} & Coverage Criteria \\ 
  An Automatic Approach for Uniquely Discovering Actionable Elements for Systematic GUI Testing in Web Applications & \cite{P268} & Test Maintenance \\ 
  Evaluation of the Choice of LLM in a Multi-Agent Solution for GUI-Test Generation & \cite{P293} & Testing Framework \\ 
  LLM-Based Labelling of Recorded Automated
GUI-Based Test Cases & \cite{P294} & Test Maintenance \\ 
  FuzzE, Development of a Fuzzing Approach for Odoo's Tours Integration Testing Plateform & \cite{P295} & Test Generation \\ 
  E2E-Loader: A Tool to Generate Performance Tests from End-to-End GUI-Level Tests
 & \cite{P298} & Test Generation \\ 
  E2e-loader: A framework to support performance testing of web applications, & \cite{P305} & Test Generation \\ 
\bottomrule
\label{tab:collab_papers}
\end{longtable}

  \bibliographystyle{elsarticle-num} 
 \bibliography{papers}



\end{document}